\documentclass[%
 reprint,
 onecolumn,
 groupedaddress,
 unsortedaddress,
 nofootinbib,
 amsmath,amssymb,
 aps,
 prd,
]{revtex4-1}

\usepackage{graphicx}% Include figure files
\usepackage{dcolumn}% Align table columns on decimal point
\usepackage{bm}% bold math
\usepackage[normalem]{ulem}
\usepackage{times}
\usepackage{color}
\usepackage{hyperref}% add hypertext capabilities
\begin{document}

\preprint{APS/123-QED}

\title{Equivalence of cosmological observables in conformally related scalar tensor theories }

\author{Fran\c cois Rondeau}
 \email{francois.rondeau@ens-paris-saclay.fr}
\author{Baojiu Li}%
 \email{baojiu.li@durham.ac.uk}
\affiliation{Institute for Computational Cosmology, Department of Physics, Durham University, Durham DH1 3LE, UK}


\begin{abstract}

Scalar tensor theories can be expressed in different frames, such as the commonly-used Einstein and Jordan frames, and it is generally accepted that cosmological observables are the same in these frames. We revisit this by making a detailed side-by-side comparison of the quantities and equations in two conformally related frames, from the actions and fully covariant field equations to the linearised equations in both real and Fourier spaces. This confirms that the  field and conservation equations are equivalent in the two frames, in the sense that we can always re-express equations in one frame using relevant transformations of variables to derive the corresponding equations in the other. We show, with both analytical derivation and a numerical example, that the line-of-sight integration to calculate CMB temperature anisotropies can be done using either Einstein frame or Jordan frame quantities, and the results are identical, provided the correct redshift is used in the Einstein frame ($1+z\neq1/a$).

\end{abstract}

\pacs{Valid PACS appear here}% PACS, the Physics and Astronomy
\maketitle

%\tableofcontents

\section{\label{sec:level1}Introduction}

The accelerated expansion of the Universe \cite{Perlmutter1999,Riess1998} observed about two decades ago is one of the most challenging questions for cosmologists and physicists today. Such an accelerated expansion cannot be explained so far in the standard framework which is built upon the standard model of particle physics and Einstein theory of General Relativity (GR), and therefore hints that new physics beyond our current knowledge might be its driving force. This makes it a very interesting and potentially very important question, and has motivated various ongoing and planned astronomical surveys, such as e{\sc boss} \cite{eboss}, {\sc des} \cite{des}, {\sc hsc} \cite{hsc}, {\sc desi} \cite{desi}, {\sc lsst} \cite{lsst}, {\sc Euclid} \cite{euclid}, 4{\sc most} \cite{4most}, {\sc wfirst} \cite{wfirst} and {\sc ska} \cite{ska}, which are designed to measure various properties of the cosmic acceleration which can in turn be used to shed light on its origin and underlying physics.

From a phenomenological point of view, the simplest possibility to explain the observations is the $\Lambda$ cold dark matter ($\Lambda$CDM) model, where a small positive cosmological constant $\Lambda$ is assumed to be accelerating the rate of the Hubble expansion. Although this model is compatible with many observations, it has suffered from theoretical difficulties, such as the fine-tuning and coincidence problems. In order to avoid these problems, various theories of dark energy \cite{DE} and modified gravity \cite{MG1,MG2} have been studied, many of which can be classified as subclasses of the so-called scalar tensor theories \cite{STT}. While GR is a tensor theory, in which the mediators of the gravitational interaction (gravitons) are excitations of the metric of the space-time, in a scalar-tensor theory, a second mediator of gravity is considered -- a scalar field $\phi$, which couples to the matter or gravitational fields (or both in certain models). The theory can then be studied in different `frames' by suitable field redefinitions. The commonly used frames include the {\em Einstein frame}, where the matter fields -- rather than the gravitational field $g_{\mu \nu}$ -- are coupled to $\phi$ and the gravity sector takes its standard form as in GR, and the {\em Jordan frame}, where the matter fields are uncoupled to $\phi$ but the gravitational equations are modified due to the coupling to $\phi$. The field equations generally look different in the two frames, and usually for certain applications in practice it is advantageous to use one over the other. 

Such a freedom of choosing to work in different frames used to be a source for debates in the community, about whether the Einstein and Jordan frames are physically equivalent to each other. The current prevailing opinion is that physics is the same in these frames, but quantities calculated in them need to be interpreted carefully to compare with each other (see, e.g., \cite{Flanagan2004,Catena2007,Faraoni2007,Veiled_GR,Chiba2013,Equivalence EF/JF} and references therein; for opposite views see, e.g., \cite{Faraoni1999,Non equivalence EF/JF}). For example, in \cite{Defelice2010} it is suggested that calculations can be done in the Einstein frame where the scalar field is a canonical field with minimal coupling to gravity (therefore simplifying the equations) but the interpretation of physical observables should be done in the Jordan frame, which is the `physical frame'. In Ref.~\cite{Catena2007} it is proposed that calculations can be done using frame-independent quantities such that the issue of how to interpret them can be circumvented naturally; \cite{Equivalence EF/JF} similarly demonstrates the equivalence of physics in the two frames by rewriting the action in terms of dimensionless variables which are frame independent; \cite{Chiba2013} finds the correspondences between variables in the two frames and show that cosmological observables, such as redshift, luminosity distance and temperature anisotropies, are frame-independent. Although these are useful works, it would also be helpful to have a detailed comparison of the linear perturbation variables (not necessarily observables) and equations in the two frames, which will provide further insight about what in a perturbed spacetime are (not) affected by a conformal transformation.

In this paper, we would like to have a closer look at the field equations in the two frames up to first order in linear perturbations. This differs from previous works in that we will write the equations -- the fully covariant Einstein and Klein Gordon equations, and their linearised versions in both real space and Fouier space -- side by side and demonstrate their equivalence. Using these equations, we will show that physical observables which are gauge invariant, such as the cosmic microwave background (CMB) temperature anisotropies, weak gravitational lensing and the integrated Sachs Wolfe effect, are the same no matter quantities in which frame are used to calculate them. The derivations will then be supplemented by a numerical example with which we show that the CMB temperature spectra calculated in the two frames are identical provided that care is taken in the Einstein frame so that integrations stop at the correct time (which is not necessarily when the scale factor $a=1$).  

% Many debates occurred about the physical equivalence of these two frames, sometimes considered as being equivalent \cite{Equivalence EF/JF}, sometimes not \cite{Non equivalence EF/JF}. One of the fundamental question is : in which frame observations and measures are done ? Is there a physical frame, where observations are carried out while the other one is just a purely mathematical transformation of the first one, or do observations can be carried out in both frames ?

% The main aim of this paper is to show that the CMB power spectrum, which is a measurable quantity, has the same expression in the Einstein and Jordan frames, and so likewise for all quantities derived from it \cite{}. This corroborates the idea that no frame is more physical than the other : whatever the observations are done in the Einstein or Jordan frames, the results remain the same.\\

This paper is organised as follows. In Section \ref{sect:math_setup}, we describe general relations between operators and mathematical quantities built from two conformally related metrics. In Section \ref{sect:scalar_tensor}, we present the physical model: a scalar-tensor theory built from the metric tensor $g_{\mu\nu}$ and a scalar field $\phi$. The Einstein equations, Klein-Gordon equations and Friedmann equations are derived independently in the two frames, and we check that any equation in one frame can always be obtained directly from its equivalent in the other frame. %This points highlights that two approaches can be followed : working in parallel and independently in the Einstein and Jordan frame, or working in a given frame and then moving to the other using the transformations law of the quantities presented in this paper. 
In Section \ref{sect:pert_eqns} we present the relations of linear perturbation variables in the two frames, and show that the linearised Einstein and matter conservation equations in the two frames are equivalent (in the sense that an equation in one frame can be derived from the same equation in the other frame using the above relations). These will be done in both real space and Fourier space which is more convenient to solve the set of coupled linear equations. In Section \ref{sect:observables} we find expressions of some of the most well-known cosmological observables and explicitly show that they are identical in both frames using the relations between linear perturbation variables in the frames; we explicitly demonstrate this by showing the CMB spectra for a specific scalar field model -- the K-mouflage model \cite{kmouflage1,kmouflage2}, which has a scalar field that is purely kinetic with a non-canonical kinetic term. 
% So far, the form of the Lagrangian density $\mathcal L_{\phi}$ of the scalar field has not been specified, and hence these results can be used for any particular Scalar-Tensor Theory models. We then choose a K-mouflage model, where $\mathcal L_{\phi}$ is purely kinetic with a non-canonical kinetic term, which is the scalar tensor model used in our numerical computations. In section IV we derive the links between several physical quantities in the Einstein and Jordan frame : fundamental quantities (masses and lengths), kinematical and dynamical quantities. Using all the results derived in the previous parts, we then show in section V that the CMB power spectrum has the same expression in the Einstein and Jordan frames. Discussions about baryonic acoustic oscillations, the integrated Sachs-Wolfe effect and weak gravitational lensing are also added, leading to the same conclusion as for the CMB power spectrum : all this observables effects have the same expressions in the Einstein and Jordan frames. 
Finally, we discuss and conclude in Section \ref{sect:discussion}. 

Throughout this work, we shall use the unit $c=1$, where $c$ is the speed of light, unless where $c$ is explicitly written. We adopt the metric sign convention $(-,+,+,+)$.

\section{The mathematical setup}
\label{sect:math_setup}

In this section we will set up the mathematical framework of this paper, by introducing notations and conventions to be used later, as well as some useful relationships between quantities in frames that are conformally related. 

Later in the paper we will use untildered and tildered quantities for the Einstein and Jordan frames respectively. For this section, however, we prefer to keep things more general and so refrain from making connections of the untildered (tildered) frame to the Einstein (Jordan) frame.

\subsection{The $3+1$ space-time decomposition}

In this paper we will focus on the analysis and predictions of observable quantities in the linear perturbation regime. The linear perturbation equations will be derived using the $3+1$ decomposition \cite[see, e.g.,][]{3+1a,3+1b}, which splits the 4D space-time into a time direction and 3D spatial slices (hypersurfaces) with constant times. The split is done with respect the 4-velocity $u^{\mu}$ of an observer. Note that as per the standard convention Greek indices run over $0,1,2,3$. Although later in the paper we will use tildered and untildered quantities for the different frames, the expressions in this subsection are mostly definitions and general for any frame, so we make all quantities un-tildered.

A projection tensor $h_{\mu \nu}$ can be defined as 
\begin{equation}
h_{\mu \nu} \equiv g_{\mu \nu} + u_{\mu}u_{\nu},
\end{equation}
which is the metric tensor of the 3D hyperspace orthogonal to $u^{\mu}$. The covariant spatial derivative $\hat{\nabla}^{\mu}$ of a general tensor field $B_{\nu... \rho}^{\sigma...\lambda}$ can then be expressed using $h_{\mu\nu}$ as
\begin{equation}
\hat{\nabla}^{\mu} B_{\nu... \rho}^{\sigma...\lambda} \equiv h^{\mu}_{\alpha} h^{\sigma}_{\beta}...h^{\lambda}_{\gamma} h^{\delta}_{\nu}...h^{\epsilon}_{\rho} \nabla^{\alpha} B_{\delta...\epsilon}^{\beta...\gamma},
\end{equation}
in which $\nabla$ denotes the full covariant derivative compatible with $g_{\mu\nu}$. Similarly, the covariant time derivative can be expressed as 
\begin{equation}
\dot{B}^{\sigma\dots\lambda}_{\nu\dots\rho} \equiv u^{\mu}\nabla_{\mu}B^{\sigma\dots\lambda}_{\nu\dots\rho}.
\end{equation}

The stress-energy tensor of matter and the scalar field can be decomposed as
\begin{equation}\label{split stress energy tensor}
T_{\mu \nu} = \rho u_{\mu} u_{\nu} + p h_{\mu \nu} + 2 q_{(\mu} u_{\nu)} + \Pi_{\mu \nu},
\end{equation}
which gives dynamical quantities including the energy density $\rho$, isotropic pressure $p$, energy flux $q_{\mu}$ and the anisotropic stress $\Pi_{\mu \nu}$. The latter two are purely spatial quantities satisfying $u^{\mu}q_{\mu}=u^{\mu}\Pi_{\mu\nu}=0$, and therefore vanish in an exact Friedman-Robertson-Walker (FRW) universe; in a perturbed FRW space-time they are linear-order quantities.

Similarly, the covariant derivative of the 4-velocity can be split as
\begin{equation}
\nabla_{\mu} u_{\nu} %= - u_{\mu} \dot u_{\nu} + \hat{\nabla}_{\mu} u_{\nu} 
= - u_{\mu} \dot u_{\nu} + \frac{1}{3} \theta h_{\mu \nu} + \sigma_{\mu \nu} + \varpi_{\mu \nu},
\end{equation}
which gives the kinematic quantities including the expansion scalar $\theta \equiv \nabla_{\alpha} u^{\alpha}$, the shear tensor $\sigma_{\mu \nu} \equiv \hat{\nabla}_{ (\mu } u_{\nu )} - \frac{1}{3} \theta h_{\mu \nu}$, the vorticity $\varpi_{\mu \nu} \equiv \hat{\nabla}_{[\mu} u_{\nu ]}$ and the 4-acceleration of the observer $w_{\mu}\equiv\dot{u}_{\mu}$. Only $\theta$ is nonzero in an exact FRW space-time, and the other three are purely spatial tensors, which are first order quantities for $w_{\mu}$ and $\sigma_{\mu\nu}$ and second order quantity for $\varpi_{\mu\nu}$ in a perturbed FRW space-time, satisfying $u^{\mu}w_{\mu}=u^{\mu}\sigma_{\mu\nu}=u^{\mu}\varpi_{\mu\nu}=0$.

Note that the metric signature used here is $(-,+,+,+)$, so that $u^{\alpha}u_{\alpha} = -1$. The Riemann tensor is defined in terms of the Christoffel symbols as 
\begin{equation}
{R^{\mu}}_{\nu \rho \sigma} = \partial_{\rho} {\Gamma^{\mu}}_{\nu \sigma} - \partial_{\sigma} {\Gamma^{\mu}}_{\nu \rho} + {\Gamma^{\mu}}_{\rho \alpha} {\Gamma^{\alpha}}_{\nu \sigma} - {\Gamma^{\mu}}_{\sigma \alpha} {\Gamma^{\alpha}}_{\nu \rho},
\end{equation} 
and the Ricci tensor and Ricci scalar are given by $R_{\mu \nu} \equiv {R^{\alpha}}_{\mu \alpha \nu}$ and $R \equiv g^{\mu \nu} R_{\mu \nu} = {R^{\mu}}_{\mu}$. %We use units in which the speed of light $c=1$.

As mentioned in the introduction, we hope that this work will be a useful reference in which linear perturbation equations in both frames are compared side by side and can be found for future work. In the literature, linear perturbation analyses are often done using the synchronous or Newtonian gauges \citep{Ma1995}. The equations in the $3+1$ formalism presented here can be re-expressed in general gauges by gauge fixing. For example, the synchronous (Newtonian) gauge corresponds to setting $w=0$ ($\sigma=0$) in our equations \citep[see, e.g.,][for an explicit mapping to those gauges]{Barreira2012}, where $w, \sigma$ are respectively the Fourier-space expressions of the scalar modes\footnote{In this work we shall only consider scalar modes of linear perturbations.} of $w_{\mu}$ and $\sigma_{\mu\nu}$ (see below).

\subsection{Mathematical quantities in conformal transformation}

%All along this paper, we will compute and compare different quantities built from two metrics 

Although the ultimate goal of this paper is to consider the equivalence of physically observable quantities in the Einstein and Jordan frames, it is useful to know how general mathematical quantities are connected in generally conformally-related frames. These relations will also be useful when we compare the equations in the two frames.

The metric tensors in two frames which later will be called the Einstein frame ($g_{\mu \nu}$) and the Jordan frame ($\tilde g_{\mu \nu}$) are related by a conformal transformation\footnote{Although we will identify these metrics to be the Einstein- and Jordan-frame ones respectively later, in this section we shall avoid doing this to keep things general. Instead, if needed we shall call them untildered and tildered frames.}
\begin{equation}\label{conformal transformation}
\tilde g_{\mu \nu} = A(\phi) \ g_{\mu \nu}, 
\end{equation}
where $A$ is a function of a scalar field $\phi$. If we denote by $g$ and $\tilde g$ the determinants of the metrics $g_{\mu \nu}$ and $\tilde g_{\mu \nu}$ respectively, we then have the useful relation
\begin{equation}\label{determinant}
\sqrt{- \tilde g} = A^2 \ \sqrt{- g}.
\end{equation}

%\subsection{Operators and Christoffel symbols}

Since the two frames are related by a conformal transformation, if we use conformal time (denoted by $\eta$) and comoving spatial coordinates (${\bf x}$), then the coordinates will be the same in the two frames. For example, in the case of exactly FRW space-times, the line elements in the two frames can be written as 
\begin{eqnarray}\label{eq:frw_metrics}
{\rm d}s^2 &=& a^2(-{\rm d}\eta^2 + \rm{d}{\bf x}^2),\nonumber\\
{\rm d}\tilde{s}^2 &=& \tilde{a}^2(-{\rm d}\eta^2 + \rm{d}{\bf x}^2).
\end{eqnarray}
Therefore, it is convenient to define the covariant derivatives using the conformal and comoving coordinates. In this case, for a general scalar field $\psi$, 
%In this paper, $\nabla_{\mu} \cdot$ means a derivative with respect to the comoving coordinates. Hence, for a given {\em dimensionless} scalar function $\psi$, %such that $\tilde \psi = \psi$, 
the covariant derivatives with lower index (which are equal to the partial derivatives% with lower index
) are the same in the two frames%\footnote{In the case of a non-dimensionless scalar field $\phi$, we have $\tilde \nabla_{\mu} \tilde \phi = \nabla_{\mu} \tilde \phi$, but $\tilde \nabla_{\mu} \tilde \phi \ne \nabla_{\mu} \phi$, as now $\phi$ is not frame-independent : $\tilde \phi \ne \phi$.} :
\begin{equation}\label{covariant derivative 1}
\tilde \nabla_{\mu} \psi = \nabla_{\mu} \psi.
\end{equation}
In the mean time, extra care needs to be taken for the covariant derivatives with upper index, as the index is not raised by the same metric in the two frames. We have $\nabla^{\mu} \psi = g^{\mu \alpha} \nabla_{\alpha} \psi$ and $\tilde \nabla^{\mu} \psi = \tilde g^{\mu \alpha} \tilde \nabla_{\alpha} \psi$, which satisfy
\begin{equation}\label{covariant derivative 2}
\tilde \nabla^{\mu} \psi = \frac{1}{A} \nabla^{\mu} \psi,
\end{equation}
where we have used $\tilde g^{\mu \nu} = \frac{1}{A} \ g^{\mu \nu}$, which is the inverse relation of Eq.~\eqref{conformal transformation}.

The covariant time derivatives in the two frames are respectively defined by $\dot \psi \equiv u^{\alpha} \nabla_{\alpha} \psi$ and $\mathring{\psi} \equiv \tilde u^{\alpha} \tilde \nabla_{\alpha} \psi$. In order to find the link between these two quantities, we have used the relation between the 4-velocities in the two frames:
\begin{equation}\label{eq:vel_relation}
\tilde{u}^{\mu} = \frac{{\rm d}\tilde{x}^{\mu}}{{\rm d}\tilde{s}} = \frac{{\rm d}{x}^{\mu}}{{\rm d}\tilde{s}} = \frac{{\rm d}{x}^{\mu}}{\sqrt{A}{\rm d}{s}} = \frac{1}{\sqrt{A}}u^{\mu},
\end{equation}
where the third equality is because ${\rm d}\tilde{s}=\sqrt{A}{\rm d}s$. Similarly,
\begin{equation}
\tilde{u}_{\mu} = \tilde{g}_{\mu\nu}\tilde{u}^{\nu} = Ag_{\mu\nu}\frac{1}{\sqrt{A}}u^{\nu}=\sqrt[]{A}u_{\mu},
\end{equation}
% Acting on a dimensionless scalar function $\psi$, they are related by :
and therefore
\begin{equation}\label{cosmic time derivatives}
\mathring{\psi} = \frac{1}{\sqrt A} \dot \psi.
\end{equation}
%which follow from the transformation law of the 4-velocity $u^{\mu}$ given in \eqref{velocity}.\\

Although Eq.~(\ref{cosmic time derivatives}) is useful in connecting the physical time derivatives in the two frames, it is more convenient to use the conformal time derivative $' \equiv {\rm d}/{\rm d}\eta$ because it is the same in both frames:
%, will also be used in what follow : the derivative with respect to the conformal time $\eta$, related to the cosmic time $t$ in the Einstein frame by $dt = a \ d\eta$ and to the cosmic time $\tilde t$ in the Jordan frame by $\tilde dt = \tilde a \ d\eta$, where $a$ and $\tilde a$ are the scale factors in the Einstein and Jordan frames. We find that this derivative with respect to the conformal time is related with derivatives with respect to the cosmic time in the Einstein and Jordan frames by:
\begin{equation}\label{cosmic and conformal time derivatives}
\psi' = \tilde{a}\mathring{\psi} = a\dot{\psi},% = \frac{1}{a} \psi' \ \ ; \ \ \mathring{\psi} = \frac{1}{\tilde a} \psi'
\end{equation}
%^As the conformal time $d \eta$ is the same in both frames, the link between the scale factors $a$ and $\tilde a$ follows directly from eq. \eqref{cosmic time derivatives} and \eqref{cosmic and conformal time derivatives} :
where $a$ and $\tilde{a}$ are the scale factors in the two frames introduced in Eqs.~\eqref{eq:frw_metrics} and they satisfy $\tilde a = a \sqrt A$.

Using the 3+1 space-time decomposition introduced in the previous subsection, % with respect to an observer's 4-velocity $u^{\mu}$, 
the covariant derivatives of $\psi$ can be split %into a time part $\dot \psi$ (along $u^{\mu}$) and a spatial part $\hat{\nabla}_{\mu} \psi$ (orthogonal to $u^{\mu}$) 
as
\begin{eqnarray}
\nabla_{\mu} \psi &=& - u_{\mu} \dot \psi + \hat{\nabla}_{\mu} \psi,\nonumber\\
\tilde \nabla_{\mu} \psi &=& - \tilde u_{\mu} \mathring{\psi} + \hat{\tilde \nabla}_{\mu}\psi,
\end{eqnarray}
in the two frames respectively, from which we obtain the following relations between the spatial derivatives:
%
%The results \eqref{covariant derivative 1}, \eqref{covariant derivative 2}, \eqref{cosmic time derivatives} and the transformation law for the 4-velocity $u^{\mu}$ given by \eqref{velocity} provide the transformation law for the spatial covariant derivatives $\hat \nabla_{\mu} \psi$ and $\hat \nabla^{\mu} \psi$ :
\begin{equation}
\hat{\tilde \nabla}_{\mu} \psi = \hat{\nabla}_{\mu} \psi \ \ ; \ \ \hat{\tilde \nabla}^{\mu} \psi = \frac{1}{A} \hat{\nabla}^{\mu} \psi.
\end{equation}
%Note that these transform in the same way as the covariant derivatives $\nabla_{\mu} \psi$ and $\nabla^{\mu} \psi$.

%In the untildered frame, $\nabla_{\mu} \cdot$ means a derivative with respect to the comoving coordinates, whereas $\nabla_{\mu} \cdot \vert_{a \chi}$ means a derivative with respect to the physical coordinates. We have :

%\begin{equation}
%\nabla_{\mu} \cdot \vert_{a \chi} = \sqrt{A} \nabla_{\mu} \cdot
%\end{equation}

%\begin{equation}
%\hat {\tilde \Box} \psi = \hat{\Box} \psi
%\end{equation}

%transformation law for the spatial metric :
%\begin{equation}\label{spatial metric}
%\tilde h_{\mu \nu} = A \ h_{\mu \nu}
%\end{equation}

Finally, starting from the expression of the Christoffel symbols:  %expressed in terms of the metric, 
\begin{equation}
\Gamma_{\mu \nu}^{\lambda} = \frac{1}{2} g^{\lambda \alpha} \left( \partial_{\mu} g_{\alpha \nu} + \partial_{\nu} g_{\mu \alpha} - \partial_{\alpha} g_{\mu \nu} \right),\nonumber
\end{equation}
in the untildered frame, and 
\begin{equation}
\tilde \Gamma_{\mu \nu}^{\lambda} = \frac{1}{2} \tilde g^{\lambda \alpha} \left( \partial_{\mu} \tilde g_{\alpha \nu} + \partial_{\nu} \tilde g_{\mu \alpha} - \partial_{\alpha} \tilde g_{\mu \nu} \right),\nonumber
\end{equation}
in the tildered frame, and using Eq.~\eqref{conformal transformation}, one can compute the link between these two quantities
\begin{equation}\label{Christoffel symbols}
\tilde \Gamma_{\mu \nu}^{\lambda} = \Gamma_{\mu \nu}^{\lambda} + \frac{1}{2A} \left( \delta_{\nu}^{\lambda} \nabla_{\mu} A
+ \delta_{\mu}^{\lambda} \nabla_{\nu} A - g_{\mu \nu} \nabla^{\lambda} A \right).
\end{equation}
This expression is useful to compute the relations between quantities involving covariant derivatives applied to general tensors. An example of its application is the computation of the link between the d'Alembertian operators in the untildered and tildered frames, respectively defined by $\Box \equiv \nabla_{\alpha} \nabla^{\alpha}$ and $\tilde \Box \equiv \tilde \nabla_{\alpha} \tilde \nabla^{\alpha}$
\begin{equation}
\tilde \Box \psi = \frac{1}{A} \Box \psi + \frac{1}{A^2} \nabla^{\alpha} \psi \nabla_{\alpha} A.
\end{equation}

%The components of the Weyl tensor in the two frames are related by :
%\begin{equation}\label{Weyl tensor}
%\tilde {C_{\mu \nu \rho}}^{\sigma} = {C_{\mu \nu \rho}}^{\sigma} \ \ ; \ \ \tilde C_{\mu \nu \rho \sigma} = A \ C_{\mu \nu \rho \sigma}
%\end{equation}
%We can use the latter relation and \eqref{determinant} and \eqref{velocity} to find the links for the electric and magnetic part of the Weyl tensor in the two frames, respectively defined by $E_{\mu \nu} \equiv C_{\mu \alpha \nu \beta} u^{\alpha} u^{\beta}$ and $H_{\mu \nu} \equiv \frac{1}{2} u^{\delta} \epsilon_{\delta \mu \alpha \beta} C^{\alpha \beta}_{\nu \gamma} u^{\gamma}$ :
%\begin{equation}
%\tilde E_{\mu \nu} = E_{\mu \nu} \ \ ; \ \ \tilde H_{\mu \nu} = H_{\mu \nu}
%\end{equation}
%and their spatial covariant derivatives :
%\begin{equation}
%\hat{\tilde \nabla}^{\alpha} \tilde E_{\mu \alpha} = \frac{1}{A} \hat{\nabla}^{\alpha} E_{\mu \alpha} \ \ ; \ \ \hat{\tilde \nabla}^{\alpha} \tilde H_{\mu \alpha} = \frac{1}{A} \hat{\nabla}^{\alpha} H_{\mu \alpha}
%\end{equation}

\subsection{Lagrangian densities and the scalar field}

In order to derive and compare the relevant equations in the two frames, we also need to know how the Lagrangian densities and a general scalar field transform in the conformal transformation.

The action built from a Lagrangian density $\mathcal L$, expressed in terms of the untildered metric, reads
\begin{equation}
S[g_{\mu \nu}] = \int{\rm d}^4x \sqrt{-g} \mathcal L,
\end{equation}
Using the transformation law for $\sqrt{-g}$ given by Eq.~\eqref{determinant}, it can be expressed in terms of the tildered metric as
\begin{equation}
S[\tilde g_{\mu \nu}] = \int{\rm d}^4x \sqrt{- \tilde g} \frac{\mathcal L}{A^2},
\end{equation}
where note that the coordinates are the same in these expressions because they are comoving coordinates.

One can then define the Lagrangian density in the tildered frame, $\tilde {\mathcal L}$, as% from the Lagrangian density in the untildered frame $\mathcal L$ by the relation \footnote{This result can be found with dimensional arguments : a Lagrangian density has the dimension of $M^4$, where $M$ is a mass. The transformation law for masses in the two frames given in the following (eq. \eqref{mass}), then leads to \eqref{Lagrangian}.} :
\begin{equation}\label{Lagrangian}
\tilde {\mathcal L} = \frac{1}{A^2} \mathcal L,
\end{equation}
% A crucial point to consider when we work with a scalar field is its physical dimension. If a scalar field $\psi$ is dimensionless, then it is frame-independent : $\tilde \psi = \psi$, but the case of a non dimensionless scalar field is more subtle. To see this, let's consider a non dimensionless scalar field $\phi$, whose Lagrangian density is purely kinetic and reads in the tildered frame :
which can be used to determine the transformation law of the scalar field between the two frames. To do this, let us have a simple example of the Lagrangian density of a scalar filed, which takes the canonic form in the tildered frame,
\begin{equation}\label{eq:sf_lag_tilder}
\tilde{\mathcal{L}}_{\tilde \phi} = - \frac{1}{2} \tilde g^{\alpha \beta} \tilde \nabla_{\alpha} \tilde \phi \tilde \nabla_{\beta} \tilde \phi.
\end{equation}
Since the transformed scalar field, $\tilde{\phi}$, is related to the scalar field $\phi$ in the untildered frame, we can write $\tilde{\nabla}_{\mu}\tilde{\phi}=(\partial\tilde{\phi}/\partial\phi)\tilde{\nabla}_{\mu}\phi$. Then, using $\tilde g^{\mu \nu} = \frac{1}{A} g^{\mu \nu}$ and $\tilde \nabla_{\mu} \phi = \nabla_{\mu} \phi$, Eq.~(\ref{eq:sf_lag_tilder}) can be rewritten as 
\begin{equation}
\tilde{\mathcal{L}}_{\tilde \phi} = - \left( \frac{\partial \tilde \phi}{\partial \phi} \right)^2 \frac{1}{2} \frac{g^{\alpha \beta}}{A} \nabla_{\alpha} \phi \nabla_{\beta} \phi.
\end{equation}
Hence, for the scalar field Lagrangian density in the untildered frame, $\mathcal{L}_{\phi}=-\frac{1}{2}g^{\alpha\beta}\nabla_{\alpha}\phi\nabla_{\beta}\phi$, the relation in Eq.~\eqref{Lagrangian} implies that the scalar fields $\phi$ and $\tilde \phi$ %expressed in both frames 
should satisfy 
\begin{equation}
\frac{\partial \tilde \phi}{\partial \phi} = \frac{1}{\sqrt A}.
\end{equation}
This in turn means that
\begin{equation}\label{derivative of Lagrangian}
\frac{\partial \tilde{\mathcal L}_{\tilde \phi}}{\partial (\tilde \nabla_{\mu} \tilde \phi)} = \frac{1}{A^{\frac{3}{2}}} \frac{\partial {\mathcal L}_{\phi}}{\partial (\nabla_{\mu} \phi)},
\end{equation}
from which and Eq.~\eqref{Christoffel symbols} we find
\begin{equation}\label{covariant derivative of derivative of Lagrangian}
\tilde \nabla_{\mu} \left[ \frac{\partial \tilde{\mathcal L}_{\tilde \phi}}{\partial (\tilde \nabla_{\mu} \tilde \phi)} \right] = \frac{1}{A^{\frac{3}{2}}} \nabla_{\mu} \left[ \frac{\partial {\mathcal L}_{\phi}}{\partial (\nabla_{\mu} \phi)} \right] + \frac{1}{2A^{\frac{3}{2}}} \frac{{\rm d}\ln A}{{\rm d}\phi}\frac{\partial\mathcal{L}_{\phi}}{\partial\left(\nabla_{\mu}\phi\right)}\nabla_{\mu}\phi.
\end{equation}
This relation can be used to transform the Klein-Gordon equation between the untildered and the tildered frames.

\section{Scalar-tensor theories}
\label{sect:scalar_tensor}

We shall consider the classical scalar-tensor theory where gravity is mediated by a scalar field in addition to the metric tensor. Let $\mathcal{M}$ be a 4-dimensional manifold representing the space-time, then $\mathcal{M}$ equipped with a metric $g_{\mu\nu}$, in which the Ricci scalar is not coupled to any scalar field, is denoted by $(\mathcal{M}, g_{\mu \nu})$ and we call this the {\em Einstein frame}. In this frame, the Einstein equations take their standard form as in GR, but matter is non-minimally coupled to the scalar field and hence free particles do not follow geodesics of the metric $g_{\mu \nu}$ but feel an additional fifth force.

The total action in the Einstein frame reads
\begin{eqnarray}\label{action EF}
S[g_{\mu \nu}] &=& \int{\rm d}^4x \sqrt{- g} \left\{\frac{1}{2} M_{\rm Pl}^2 R + \mathcal{L}_{\phi} [\phi, (\nabla \phi)^2] \right \} + S_m [A(\phi), g_{\mu \nu}],
\end{eqnarray}
where $M_{\rm Pl}$ is the reduced Planck mass defined by $M_{\rm Pl}^{-2} = 8 \pi G$, $G$ is Newton's constant, $\mathcal{L}_{\phi}$ is the Lagrangian density of the scalar field $\phi$, $R=g_{\mu\nu}R^{\mu\nu}$ is the Ricci scalar, $R_{\mu\nu}$ the Ricci tensor, and $S_m[A(\phi), g_{\mu\nu}]$ is the matter action given by 
\begin{equation}
S_m[A(\phi), g_{\mu \nu}] = \sum_i \int{\rm d}^4x \sqrt{-g} \mathcal{L}_m(\psi_m^{(i)}, A(\phi), g_{\mu \nu}),
\end{equation}
where $A(\phi)$ is an algebraic function of the scalar field $\phi$, $\psi_m^{(i)}$ denotes the $i$th species of matter fields, and the summation is over all matter species.

We can introduce another, tildered, metric $\tilde{g}_{\mu\nu}$ through $\tilde{g}_{\mu\nu}=A(\phi)g_{\mu\nu}$, and 
% In the Einstein frame, the Ricci scalar is minimally coupled to the scalar field $\phi$, but the matter is non-minimally coupled to the scalar field. One can remove this coupling between the matter and the scalar field by performing a conformal transformation of the metric :
% \begin{equation}
% \tilde g_{\mu \nu} = A(\phi) \ g_{\mu \nu} 
% \end{equation}
% which implies
the relation between the Ricci scalars of these two metrics can be straightforwardly found by using the transformations of the Christoffel symbols given in Eq.~(\ref{Christoffel symbols}):
\begin{equation}\label{Ricci scalar}
R = A \tilde R + 3 \ \tilde{\Box} A - \frac{9}{2} \frac{(\tilde \nabla A)^2}{A}.
\end{equation}
Using this relation and the transformations of the metric tensors, we can re-express the Einstein-frame action Eq.~\eqref{action EF} in terms of the tildered metric $\tilde{g}_{\mu\nu}$ and a redefined scalar field $\tilde{\phi}$:
%\begin{equation}\label{Ricci scalar}
%\tilde R = \frac{1}{A} \left(R - 3 \frac{\Box A}{A} + \frac{3}{2} \frac{(\nabla A)^2}{A^2} \right)
%\end{equation}
%
%Starting from \eqref{action EF} and using \eqref{conformal transformation}, \eqref{determinant} and \eqref{Ricci scalar}, one can compute the action expressed in terms of the metric $\tilde g_{\mu \nu}$, up to a boundary term :
\begin{widetext}
\begin{equation}\label{action JF}
S[\tilde g_{\mu \nu}] = \int{\rm d}^4x \sqrt{- \tilde g} \left\{ \frac{1}{2} M_{\rm Pl}^2 \ \frac{1}{A} \tilde R + \frac{3}{4} M_{\rm Pl}^2 \frac{(\tilde \nabla A)^2}{A^3} + \tilde{\mathcal{L}}_{\tilde \phi} (\tilde \phi, (\tilde \nabla \tilde \phi)^2) \right\} + S_m [\tilde g_{\mu \nu}],
\end{equation}
\end{widetext}
where we have dropped a boundary term which does not contribute to the dynamics of the theory, and the matter action expressed in terms of the tildered metric $\tilde g_{\mu \nu}$ is given by 
\begin{equation}
S_m[\tilde g_{\mu \nu}] = \sum_i \int{\rm d}^4x \sqrt{- \tilde g} \tilde {\mathcal{L}}_m(\tilde \psi_m^{(i)}, \tilde g_{\mu \nu}).
\end{equation}

The manifold $\mathcal{M}$ endowed with the metric $\tilde g_{\mu\nu}$, where the metric now is non-minimally coupled to the scalar field, is denoted by $(\mathcal{M}, \tilde{g}_{\mu\nu})$ and called the {\em Jordan frame}. In this frame, matter is minimally coupled to the scalar field, and free particles follow geodesics of the metric $\tilde g_{\mu \nu}$, just as in General Relativity. On the other hand, the Einstein equations are different from those in GR, highlighting the physical result that gravity is now mediated not just by the massless gravitons, but also by the scalar field.

From here on we shall use tildes for Jordan-frame quantities while their Einstein-frame counterparts are untildered.

\subsection{Einstein's equations in the two frames}

Varying the action Eq.~\eqref{action EF} with respect to the metric $g_{\mu \nu}$ gives the Einstein equations in the Einstein frame:
\begin{equation}\label{Einstein eq EF}
R_{\mu \nu} - \frac{1}{2} g_{\mu \nu} R = M_{\rm Pl}^{-2} \left(T_{\mu \nu}^{(m)} + T_{\mu \nu}^{(\phi)} \right),
\end{equation}
where the stress energy tensors $T_{\mu \nu}^{(i)}$, $i = m$ or $\phi$, are defined by $T_{\mu \nu}^{(i)} \equiv - \frac{2}{\sqrt{- g}} \frac{\delta [\sqrt{- g} \mathcal L^{(i)}]}{\delta g^{\mu \nu}}$.

Similarly, varying the action Eq.~\eqref{action JF} w.r.t.~the metric $\tilde g_{\mu \nu}$ gives the Einstein equations in the Jordan frame:
\begin{widetext}
\begin{equation}\label{Einstein eq JF}
\tilde R_{\mu \nu} - \frac{1}{2} \tilde g_{\mu \nu} \tilde R = M_{\rm Pl}^{-2} \left( A \tilde T_{\mu \nu}^{(m)} + A \tilde T_{\mu \nu}^{(\tilde \phi)} \right) \\ + \tilde g_{\mu \nu} \frac{\tilde \Box A}{A} - \frac{\tilde \nabla_{\mu} \tilde \nabla_{\nu} A}{A} + \frac{1}{2} \frac{\tilde \nabla_{\mu} A \tilde \nabla_{\nu} A}{A^2} - \frac{5}{4} \tilde g_{\mu \nu} \frac{(\tilde \nabla A)^2}{A^2},
\end{equation}
\end{widetext}
where the stress energy tensors $\tilde T_{\mu \nu}^{(i)}$, $i = m$ or $\tilde \phi$, are defined by $\tilde T_{\mu \nu}^{(i)} \equiv - \frac{2}{\sqrt{- \tilde g}} \frac{\delta [\sqrt{- \tilde g} \tilde{\mathcal L}^{(i)}]}{\delta \tilde g^{\mu \nu}}$.\\

Using the links between operators in the two frames given in Section \ref{sect:math_setup}, and the relations between the Ricci tensors and scalars in both frames, which are given by
\begin{equation}
\tilde R_{\mu \nu} = R_{\mu \nu} - \frac{\nabla_{\mu} \nabla_{\nu} A}{A} - \frac{g_{\mu \nu}}{2} \frac{\Box A}{A} + \frac{3}{2} \frac{\nabla_{\mu} A \nabla_{\nu} A}{A^2},
\end{equation}
and
\begin{equation}\label{eq:R-JF-EF}
\tilde R = \frac{1}{A} \left(R - 3 \frac{\Box A}{A} + \frac{3}{2} \frac{(\nabla A)^2}{A^2} \right),
\end{equation}
one can check that Eq.~\eqref{Einstein eq EF} is equivalent to Eq.~\eqref{Einstein eq JF}.

From Eq.~\eqref{Einstein eq JF}, we can define an effective stress-energy tensor $\tilde{\mathbf{T}}_{\mu \nu}^{(\tilde \phi)}$ for the scalar field in the Jordan frame:
\begin{widetext}
\begin{equation}\label{eq:Tmn_eff}
\tilde M_{\rm Pl}^{-2} \tilde{\mathbf{T}}_{\mu \nu}^{(\tilde \phi)} \equiv \tilde M_{\rm Pl}^{-2} \tilde T_{\mu \nu}^{(\tilde \phi)} \\ + \tilde g_{\mu \nu} \frac{\tilde \Box A}{A} - \frac{\tilde \nabla_{\mu} \tilde \nabla_{\nu} A}{A} + \frac{1}{2} \frac{\tilde \nabla_{\mu} A \tilde \nabla_{\nu} A}{A^2} - \frac{5}{4} \tilde g_{\mu \nu} \frac{(\tilde \nabla A)^2}{A^2},
\end{equation}
\end{widetext}
in which we have defined the reduced Planck mass in the Jordan frame as $\tilde{M}_{\rm Pl}=A^{-1/2}M_{\rm Pl}$, and compute the corresponding dynamical quantities using 
\begin{eqnarray}\label{eq:Tmn_def}
\tilde{\boldsymbol{\rho}}^{(\tilde \phi)} &=& \tilde{\mathbf{T}}_{\mu \nu}^{(\tilde \phi)} \tilde u^{\mu} \tilde u^{\nu},\nonumber\\ 
\tilde{\boldsymbol{p}}^{(\tilde \phi)} 
&=& \frac{1}{3} \tilde{\mathbf{T}}_{\mu \nu}^{(\tilde \phi)} \tilde h^{\mu \nu},\nonumber\\
\tilde{\boldsymbol{q}}^{\mu}_{(\tilde \phi)} &=& - \tilde{\mathbf{T}}_{\alpha \beta}^{(\tilde \phi)} \tilde u^{\alpha} \tilde h^{\beta \mu},\nonumber\\
\tilde{\boldsymbol{\Pi}}_{\mu \nu}^{(\tilde \phi)} &=& \tilde h_{\mu}^{\alpha} \tilde h_{\nu}^{\beta} \tilde{\mathbf{T}}_{\alpha \beta}^{(\tilde \phi)} - \tilde{\boldsymbol{p}}^{(\tilde \phi)} \tilde h_{\mu \nu}.
\end{eqnarray}
Note that the reduced Planck mass is related to Newton's constant, and is a fundamental constant in the Einstein frame. In the Jordan frame, however, it depends on the scalar field. Up to linear order, Eqs.~\eqref{eq:Tmn_eff} and \eqref{eq:Tmn_def} give
\begin{eqnarray}\label{scalar field energy density}
\tilde M_{\rm Pl}^{-2} \tilde{\boldsymbol{\rho}}^{(\tilde \phi)} &=& \tilde M_{\rm Pl}^{-2} \tilde \rho^{(\tilde \phi)} - \frac{\hat{\tilde \Box}A}{A} + \tilde \theta \frac{\mathring A}{A} - \frac{3}{4} \frac{{\mathring A}^2}{A^2},\\
\label{scalar field pressure} \tilde M_{\rm Pl}^{-2} \tilde{\boldsymbol{p}}^{(\tilde \phi)} &=& \tilde M_{\rm Pl}^{-2} \tilde p^{(\tilde \phi)} + \frac{2}{3} \frac{\hat{\tilde \Box}A}{A} - \frac{\mathring{\mathring A}}{A} - \frac{2}{3} \tilde \theta \frac{\mathring A}{A} + \frac{5}{4} \frac{{\mathring A}^2}{A^2},\\
\label{scalar field energy flux} \tilde M_{\rm Pl}^{-2} \tilde{\boldsymbol{q}}^{\mu}_{(\tilde \phi)} &=& \tilde M_{\rm Pl}^{-2} \tilde q^{\mu}_{(\tilde \phi)} - \frac{\mathring A}{2 A^2} \hat{\tilde \nabla}^{\mu} A + \frac{\hat{\tilde \nabla}^{\mu} \mathring A}{A} - \frac{\tilde{\theta}}{3}\frac{\hat{\tilde \nabla}^{\mu} A}{A},~~~~~~\\
\label{scalar field anisotropic stress}
\tilde M_{\rm Pl}^{-2} \tilde{\boldsymbol{\Pi}}_{\mu \nu}^{(\tilde \phi)} &=& \tilde M_{\rm Pl}^{-2} \tilde {\Pi}_{\mu \nu}^{(\tilde \phi)} - \frac{1}{A} \left( \hat{\tilde \nabla}_{\langle \mu} \hat{\tilde \nabla}_{\nu\rangle} A - \mathring{A}\tilde{\sigma}_{\mu\nu}\right).
\end{eqnarray}

At the background level, $\hat{\tilde \Box}A = 0$ and $\tilde \theta = 3 \frac{\mathring{\tilde a}}{\tilde a} = 3 \frac{{\tilde a}'}{{\tilde a}^2}$, and using $\mathring{A}=\frac{1}{\tilde a} A'$, one can rewrite Eq.~\eqref{scalar field energy density} as
\begin{equation}
\tilde M_{\rm Pl}^{-2} \tilde{\boldsymbol{\rho}}^{(\tilde \phi)} = \tilde M_{\rm Pl}^{-2} \tilde \rho^{(\tilde \phi)} + 3 \frac{{\tilde a}'}{{\tilde a}^3} \frac{A'}{A} - \frac{3}{4} \frac{{A'}^2}{{\tilde a}^2 A^2}.
\end{equation}

% Using \eqref{scalar field energy density}, \eqref{expansion scalar}, \eqref{energy density}, $\tilde a = a \sqrt A$ and $A M_{\rm Pl}^{-2} = \tilde M_{\rm Pl}^{-2}$, one can then check that the Friedmann equation in the Einstein frame 

Using this equation, it can be easily checked that the Friedmann equations in the Einstein and Jordan frames,
\begin{eqnarray}
3 \left(\frac{\dot a}{a}\right)^2 &=& M_{\rm Pl}^{-2} \left( \rho^{(m)} + \rho^{(\phi)} \right),\\
3 \left(\frac{\mathring{\tilde a}}{\tilde a}\right)^2 &=& \tilde M_{\rm Pl}^{-2} \left( \tilde \rho^{(m)} + \tilde{\boldsymbol{\rho}}^{(\tilde \phi)} \right),
\end{eqnarray}
are equivalent to each other\footnote{Throughout this paper, when we say that equations in the two frames are equivalent, we mean that we can start from the equation in one frame and derive the corresponding equation in the other frame using transformation laws of variables in these frames.}. For this check we also used the transformation of the density field, $\tilde{\rho}=A^{-2}\rho$, in the two frames, which will be discussed below. This is a consequence of the equivalence between the Einstein equations themselves in the two frames, as we checked above; this equivalence originates from the fact that the equations in the two frames were derived from the same action and thus should contain the same physics despite being expressed differently.

\subsection{The Klein-Gordon equations in the two frames}

Varying the Einstein frame action Eq.~\eqref{action EF} with respect to the scalar field $\phi$ leads to the equation of motion for $\phi$ (the Klein-Gordon equation) in the Einstein frame:
\begin{equation}\label{KG eq EF}
\nabla_{\mu} \left[ \frac{\partial \mathcal{L}_{\phi} (\phi, (\nabla \phi)^2)}{\partial(\nabla_{\mu} \phi)} \right] = \frac{\partial \mathcal{L}_{\phi}}{\partial \phi} + \frac{1}{2} \frac{{\rm d}\ln A}{{\rm d}\phi} T^{(m)},
\end{equation}
%The energy momentum tensor of the $i$th species of matter is defined by $T_{\mu \nu}^{(i)} \equiv - \frac{2}{\sqrt{-g}} \frac{\delta [\sqrt{-g} \mathcal{L}_m^{(i)}]}{\delta g^{\mu \nu}}$, $T^{(i)}$ is its trace : 
where $T^{(i)} \equiv T_{\mu \nu}^{(i)} g^{\mu \nu}$ is the trace of the total stress-energy tensor of matter for the $i$th matter species, and $^{(m)}$ means that the equation only depends on $T^{(i)}$ for all other matter species than the scalar field.% $T^{(m)} \equiv \sum_{i \ne \varphi} T^{(i)}$.\\

Similarly, varying the Jordan frame action Eq.~\eqref{action JF} with respect to the scalar field $\tilde \phi$ gives the equation of motion for $\tilde \phi$ in the Jordan frame:
\begin{widetext}
\begin{equation}\label{KG eq JF 1}
\frac{3}{2} M_{\rm Pl}^2 \left(\frac{{\rm d}A}{{\rm d}\tilde \phi}\right)^2 \frac{\tilde \Box \tilde \phi}{A^3} + \tilde{\nabla}_{\mu} \left[ \frac{\partial \tilde{\mathcal{L}}_{\tilde \phi} (\tilde \phi, (\tilde \nabla \tilde \phi)^2)}{\partial (\tilde{\nabla}_{\mu} \tilde \phi)} \right] = - \frac{1}{2} M_{\rm Pl}^2 \frac{1}{A^2} \frac{{\rm d}A}{{\rm d}\tilde \phi} \tilde R - \frac{3}{2} M_{\rm Pl}^2 \frac{{\rm d}A}{{\rm d}\tilde \phi} \frac{{\rm d}^2 A}{{\rm d}\tilde \phi^2} \frac{(\tilde \nabla \tilde \phi)^2}{A^3} + \frac{9}{4} M_{\rm Pl}^2 \left( \frac{{\rm d}A}{{\rm d}\tilde \phi} \right)^3 \frac{(\tilde \nabla \tilde \phi)^2}{A^4} + \frac{\partial \tilde{\mathcal{L}}_{\tilde \phi}}{\partial \tilde \phi}.
\end{equation}
\end{widetext}
This equation can be simplified by using the trace of Eq.~\eqref{Einstein eq JF} to replace $\tilde R$ in Eq.~\eqref{KG eq JF 1} with the trace of the stress energy tensor. Doing this leads to
\begin{equation}\label{KG eq JF 2}
\tilde \nabla_{\mu} \left[ \frac{\partial \tilde{\mathcal L}_{\tilde \phi} (\tilde \phi, (\tilde \nabla \tilde \phi)^2)}{\partial (\tilde \nabla_{\mu} \tilde \phi)} \right] = \frac{\partial \tilde {\mathcal L}_{\tilde \phi}}{\partial \tilde \phi} \\ + \frac{1}{2} \frac{{\rm d}\ln A}{{\rm d}\tilde \phi} \left(\tilde T^{(m)} + \tilde T^{(\tilde \phi)} \right).
\end{equation}
As with the case of the Einstein equations, it can be verified that the Klein-Gordon equations in the two frames, despite having different forms, are mathematically equivalent to each other. An explicit check for general scalar field Lagrangians is presented in Appendix \ref{sect:append_b}.

% Following \cite{K-mouflage}, we will work in the particular case of a K-mouflage model, where the scalar field Lagrangian is purely kinetic :

% \begin{equation}
% \mathcal{L}_{\phi}(\phi) = - M^4 K(\sigma)
% \end{equation}

% with

% \begin{equation}
% \sigma \equiv \frac{1}{2M^4} \nabla_{\mu} \phi \nabla^{\mu} \phi
% \end{equation}

% Let us notice that $\sigma$ and hence $K(\sigma)$ are frame-independent, which can be easily checked using \eqref{mass}, \eqref{covariant derivative 1} and \eqref{covariant derivative 2}.\\ 

% In this model, the Klein-Gordon equations in the Einstein and Jordan frames \eqref{KG eq EF} and \eqref{KG eq JF 2} respectively reduce to :

% \begin{equation}\label{KG eq EF K-mouflage}
% \nabla_{\mu} \left[ \partial_{\sigma} K(\sigma) \nabla^{\mu} \phi \right] = - \frac{1}{2} \frac{d \ln A}{d \phi} T^{(m)}
% \end{equation}

% and

% \begin{equation}\label{KG eq JF K-mouflage}
% \tilde \nabla_{\mu} \left[ \partial_{\sigma} K(\sigma) \tilde \nabla^{\mu} \tilde \phi \right] = - \frac{1}{2} \frac{d \ln A}{d \tilde \phi} \left( \tilde T^{(m)} + \tilde T^{(\tilde \phi)} \right)
% \end{equation}

%The following work focuses on expressing different quantities in this two frames and showing that physical quantities, ie quantities that can be measured by an observer, are the same in the two frames.

\section{Perturbation equations in the two frames}
\label{sect:pert_eqns}

In Section~\ref{sect:math_setup} we have given the relations between quantities in two general conformally-related frames. In Sect.~\ref{sect:scalar_tensor} we briefly checked that the fully covariant Einstein and Klein-Gordon equations are equivalent in the two frames, because they are derived from a same action. In order to demonstrate that physical observables are equivalent in the two frames, we next need to have a thorough look at the individual perturbation variables that are relevant to those observables. The aim of this section is to find the transformation laws of these perturbation variables and use them to show that the perturbed field and conservation equations in the two frames contain the same physics.

%The Newtonian gravitational constant is also rescaled in the Jordan frame :
%\begin{equation}\label{Newtonian constant}
%\tilde G = A \ G
%\end{equation}
%In order to see this explicitly, one can use the expression of Planck's length $l_p = \sqrt{\frac{G \hbar}{c^3}}$ and use \eqref{length} to obtain directly the result \eqref{Newtonian constant}.

\subsection{Quantities and equations in real space}

We start with the various perturbation quantities and their equations in real space, and move to $k$ (or Fourier) space in the next subsection.

\subsubsection{Kinematic quantities}

First look at the transformations of the kinematic quantities that are related to the curvature of the space-time -- the expansion scalar $\theta$, shear $\sigma_{\mu\nu}$, vorticity $\varpi_{\mu\nu}$ and 4-acceleration $w_\mu=\dot{u}_{\mu}$.

% Starting from the definition of the 4-velocity $\tilde u^{\mu} = \frac{dx^{\mu}}{\mid \tilde ds \mid}$ and using \eqref{length}, we find the relation of the 4-velocities in the two frames :
We already found above that the 4-velocities in the Einstein and Jordan frames are related by $
\tilde u^{\mu} = \frac{1}{\sqrt A} u^{\mu}, \tilde u_{\mu} = \sqrt{A} u_{\mu}$.
The %covariant and contravariant components of the 4-velocity are rescaled in the tildered frame, but the
norm of the 4-velocity is the same in the two frames -- $\tilde u^{\mu} \tilde u_{\mu} = u^{\mu} u_{\mu}=-1=-c^2$. Notice that the conformal transformation does not change the speed of light.

%This corroborates the idea that measurable quantities remain the same in the two frames.\\

Using the definitions for the time derivatives $\dot \psi \equiv u^{\alpha} \nabla_{\alpha} \psi$ and $\mathring{\psi} \equiv \tilde u^{\alpha} \tilde \nabla_{\alpha} \psi$, where $\psi$ now is a simplified notation of a generic tensor, and the relation between the Christoffel symbols Eq.~\eqref{Christoffel symbols}, one finds the 4-acceleration in the Jordan frame in terms of the one in the Einstein frame
\begin{eqnarray}\label{acceleration}
\tilde{w}^{\mu} &=& \mathring{\tilde u}^{\mu}\ = \frac{1}{A} \dot u^{\mu} + \frac{1}{2} \frac{\hat{\nabla}^{\mu} A}{A^2},\nonumber\\ \tilde{w}_{\mu} &=& \mathring{\tilde u}_{\mu}\ =\ \dot{u}_{\mu} + \frac{1}{2} \frac{\hat{\nabla}_{\mu} A}{A}.
\end{eqnarray}
This reflects the fact that the forces felt by particles in the two frames are not the same: in the Einstein frame there is an additional fifth force, which depends on the spatial gradient of $A$.

%We note that if the acceleration vanish in the Einstein frame, it does no longer vanish in the Jordan frame : an observer following a geodesic in the Einstein frame will no longer follow a geodesic in the Jordan frame, and thus feels a fifth force. However, this force is not a physical force in the sense that it can be erased by a change of metric, in the same sense that an inertial force in Newtonian gravity can be erased by a change of coordinates.\\

%This point should be taken into account in numerical computations, where the gauge should be the same in the two frames in order to compare the results.\\

%\begin{equation}\label{spatial derivative of velocity}
%\hat{\tilde \nabla}_{\mu} \tilde u_{\nu} = \sqrt{A} \hat{\nabla}_{\mu} u_{\nu} + \frac{1}{2} h_{\mu \nu} \frac{\dot A}{\sqrt A}
%\end{equation}

Using Eqs.~\eqref{Christoffel symbols} and \eqref{eq:vel_relation}, we can compute the relation between the expansion scalars in the Einstein frame and the Jordan frame, defined by $\theta \equiv \nabla_{\alpha} u^{\alpha}$ and $\tilde \theta \equiv \tilde \nabla_{\alpha} \tilde u^{\alpha}$ respectively:
\begin{equation}\label{expansion scalar}
\tilde \theta = \frac{1}{A^{\frac{1}{2}}}\theta + \frac{3}{2} \frac{\dot A}{A^{\frac{3}{2}}}.
\end{equation}
The second term comes from the fact that in the Jordan frame the scale factor is given by $\tilde{a}=\sqrt{A}a$, and an overall factor of $A^{-\frac{1}{2}}$ comes from the fact that the time derivative in $\tilde{\theta}$ ($\theta$) is with respect to $\tilde{a}{\rm d}\eta$ ($a{\rm d}\eta$): expressed in conformal time, this becomes
\begin{eqnarray}
\frac{\tilde{a}'}{\tilde{a}} &=& \frac{a'}{a}+\frac{A'}{2A}.
\end{eqnarray}
It is straightforward to find the link between the time derivatives of the expansion scalars in both frames by using Eqs.~\eqref{expansion scalar} and \eqref{cosmic time derivatives}:
\begin{equation}\label{time derivative of expansion scalar}
\mathring{\tilde \theta} = \frac{1}{A} \dot \theta - \frac{1}{2} \theta \frac{\dot A}{A^2} - \frac{9}{4} \frac{\dot A^2}{A^3} + \frac{3}{2} \frac{\ddot A}{A^2}.
\end{equation}
We can similarly find the relations between the spatial gradients of the expansion scalar, $Z_{\mu} \equiv \hat{\nabla}_{\mu} \theta$ in the Einstein frame and $\tilde{Z}_{\mu} \equiv \hat{\tilde \nabla}_{\mu} \tilde \theta$ in the Jordan frame. Using Eq.~\eqref{expansion scalar} and \eqref{time derivative of expansion scalar}, this is found to be
\begin{equation}\label{comoving spatial derivative of expansion scalar}
\tilde{Z}_{\mu} = \frac{1}{A^{\frac{1}{2}}}{Z}_{\mu} - \frac{1}{2} \theta \frac{\hat{\nabla}_{\mu} A}{A^{\frac{3}{2}}} + \frac{3}{2} \frac{\hat{\nabla}_{\mu} \dot A}{A^{\frac{3}{2}}} - \frac{9}{4} \frac{\dot A}{A^{\frac{5}{2}}} \hat{\nabla}_{\mu} A.
\end{equation}
$Z_{\mu}$ ($\tilde{Z}_{\mu}$) represents the spatial perturbation of local expansion rate.
 
The relations between the shear and vorticity tensors in the two frames can be similarly obtained as
\begin{eqnarray}\label{shear tensor}
\tilde \sigma_{\mu}^{\nu} &=& \frac{1}{A^{\frac{1}{2}}} \sigma_{\mu}^{\nu}; \ \ \tilde{\sigma}_{\mu\nu} = A^{\frac{1}{2}}\sigma_{\mu\nu}, \\
\label{vorticity tensor}
\tilde \varpi_{\mu}^{\nu} &=& \frac{1}{A^{\frac{1}{2}}} \varpi_{\mu}^{\nu}; \ \ \tilde{\varpi}_{\mu\nu} = A^{\frac{1}{2}}\varpi_{\mu\nu},
\end{eqnarray}
The relation Eq.~\eqref{shear tensor} between the shear tensors in both frames is linear; this means that if the shear tensor $\sigma_{\mu}^{\nu}$ in the Einstein frame vanishes, then the same happens for the shear tensor $\tilde \sigma_{\mu}^{\nu}$ in the Jordan frame. This relation is expected as $A$ is the conformal factor, which affects the volume and therefore the expansion rate, but not the shear.

These transformation laws of the kinematic quantities have a role to play when comparing certain gauges in the two frames. When solving the linear perturbation equations in GR, not all degrees of freedom (dofs) are physical; the non-physical dofs do not affect the physical results and can be fixed by choosing to work in some gauge. The choice of gauge is not unique and different choices usually have advantages and disadvantages. Some commonly used gauges are
% Working with the 3+1 decomposition of space-time requires to choose an observer's 4-velocity $u^{\mu}$, with respect to which the space-time is split. At the background level, the choose of $u^{\mu}$ is natural : $u^{\mu}$ is the 4-velocity of a fundamental observer, that is an observer with constant spatial comoving coordinates. At the perturbed level however, this choice is no longer unique. All 4-velocities $u^{\mu}$ defined such that they reduce to the 4-velocity of a fundamental observer in the unperturbed limit are  acceptable. The choice of a particular $u^{\mu}$ at the perturbed level is called a choice of gauge. We can list some commonly used gauges:
\begin{itemize}
\item The synchronous gauge, where the 4-acceleration of the observer $\dot{u}^{\mu}$ vanishes.
\item The Newtonian gauge, where the shear tensor $\sigma_{\mu}^{\nu}$ vanishes.
\item The energy frame, where the energy flux $q_{\mu}$ vanishes.
%\item The Cold Dark Matter frame, where the Cold Dark Matter velocity vanishes.
\end{itemize}
It is straightforward to see from Eq.~\eqref{acceleration} that the synchronous gauge is not preserved by a change of frame. For instance, if we choose to work with the synchronous gauge in the Einstein frame by setting $\dot u^{\mu} = 0$, then in the Jordan frame $\mathring{\tilde u}^{\mu}$ generally does not vanish. The Newtonian gauge, on the other hand, is the same in both frames due to Eq.~\eqref{shear tensor}. Later we will see that the same applies for the energy frame.

% Among the four gauges described above, the synchronous gauge is the only one which is not preserved by a change of frame. The CDM frame is preserved, which directly follows from \eqref{velocity}, as well as the Newtonian gauge and the energy frame, as we will see in the following.\\

\subsubsection{Dynamical quantities}

Next we turn to the relations between the dynamical quantities in the two frames, starting from the definition of the stress-energy tensor
\begin{equation}
T_{\mu \nu}^{(i)} \equiv - \frac{2}{\sqrt{-g}} \frac{\delta \left[\sqrt{-g} \mathcal L^{(i)} \right]}{\delta g^{\mu \nu}}.\nonumber
\end{equation}
Using Eqs.~\eqref{conformal transformation}, \eqref{determinant}, and the fact that the term in brackets is frame-independent, one finds the following relations
\begin{equation}\label{stress energy tensor}
\tilde{T}_{\mu \nu}^{(i)} = \frac{1}{A} T_{\mu \nu}^{(i)}; \ \ \tilde{T}_{\nu}^{\mu (i)} = \frac{1}{A^2} T_{\nu}^{\mu (i)}; \ \ \tilde{T}^{\mu \nu}_{(i)} = \frac{1}{A^3} T^{\mu \nu}_{(i)},
\end{equation}
where the superscript $^{(i)}$ indicates that these relations are valid for any species.

Using the decomposition Eq.~\eqref{split stress energy tensor} and the relations Eq.~\eqref{stress energy tensor}, we find that the energy density, isotropic pressure, energy flux and anisotropic stress in the two frames are related by
\begin{eqnarray}
\label{energy density}
\tilde \rho^{(i)} &=& \frac{1}{A^2} \rho^{(i)},\\
\label{pressure}
\tilde p^{(i)} &=& \frac{1}{A^2} p^{(i)},\\
\label{energy flux}
\tilde q^{(i)}_{\mu} &=& \frac{1}{A^{\frac{3}{2}}} q^{(i)}_{\mu}; \ \ \tilde q_{(i)}^{\mu} = \frac{1}{A^{\frac{5}{2}}} q_{(i)}^{\mu},\\
\label{anisotropic stress}
\tilde \Pi^{\nu (i)}_{\mu} &=& \frac{1}{A^2} \Pi^{\nu (i)}_{\mu}; \ \ \tilde{\Pi}^{(i)}_{\mu\nu} = \frac{1}{A}\Pi^{(i)}_{\mu\nu}.
\end{eqnarray}
Note that the relation in Eq.~\eqref{energy flux} between the energy fluxes in the Einstein and Jordan frames is linear, which confirms that the energy gauge is preserved by a conformal transformation. %change of frame, exactly like the CDM frame and the Newtonian gauge, and contrarily to the synchronous gauge.\\

One can intuitively understand the relation between the energy densities in the two frames given in Eq.~(\ref{energy density}). Since $\tilde g_{\mu \nu} = A \ g_{\mu \nu}$, the relation between infinitesimal lengths in the two frames is given as ${\rm d}\tilde{s} = \sqrt{A}{\rm d}s$. %At the background level, $A$ is a function of time only. 
Hence, if one considers a length $l$ in a spatial hypersurface of constant time $t$, it can be easily found that $\tilde l = \sqrt{A}l$. Meanwhile, the relation between masses in the two frames can be found by considering the action of a point particle of mass $\tilde{m}$ expressed in terms of the tildered metric
\begin{equation}
S_m[\tilde g_{\mu \nu}] = - \tilde m \int \sqrt{- \tilde g_{\mu \nu}{\rm d}x^{\mu}{\rm d}x^{\nu}}.
\end{equation}
Using $\tilde g_{\mu \nu} = A \ g_{\mu \nu}$, the same action can be expressed in terms of the untildered metric as
\begin{equation}
S_m[g_{\mu \nu}] = - \tilde m \int \sqrt A \sqrt{- g_{\mu \nu} {\rm d}x^{\mu} {\rm d}x^{\nu}},
\end{equation}
which implies that
\begin{equation}\label{mass}
\tilde m = \frac{1}{\sqrt A}m.
\end{equation}
Therefore, in this case the density in the Jordan frame is given by $\tilde{\rho}=\tilde{m}\tilde{l}^{-3}=A^{-2}ml^{-3}=A^{-2}\rho$.

From the above equations, we can also find the relations between the time and spatial derivatives of the energy density in the two frames, which is useful for the following sections:
\begin{eqnarray}\label{time derivative of energy density}
\mathring{\tilde \rho} &=& \frac{1}{A^{\frac{5}{2}}} \dot{\rho} - 2\frac{\dot A}{A^{\frac{7}{2}}}\rho,\\
\label{spatial derivative of energy density}
\hat{\tilde \nabla}_{\mu} \tilde \rho &=& \frac{1}{A^2} \hat{\nabla}_{\mu} \rho - 2 \rho \frac{\hat{\nabla}_{\mu} A}{A^3}.
\end{eqnarray}

% \begin{equation}\label{time derivative of energy flux}
% \mathring{\tilde q}_{\mu} = \frac{1}{A^2} \dot q_{\mu} - 2 q_{\mu} \frac{\dot A}{A^3}	
% \end{equation}

% \begin{equation}\label{spatial derivative of energy flux}
% \hat{\tilde \nabla}_{\alpha} \tilde q^{\alpha} = \frac{1}{A^{\frac{5}{2}}} \hat{\nabla}_{\alpha} q^{\alpha}
% \end{equation}

% \begin{equation}\label{spatial derivative of spatial derivative of energy flux}
% \hat{\tilde \nabla}_{\mu}  \hat{\tilde \nabla}_{\alpha} \tilde q^{\alpha} = \frac{1}{A^{\frac{5}{2}}} \hat{\nabla}_{\mu} \hat{\nabla}_{\alpha} q^{\alpha}
% \end{equation}

% \begin{equation}\label{spatial derivative of anisotropic stress}
% \hat{\tilde \nabla}_{\alpha} \tilde \Pi_{\mu}^{\alpha} = \frac{1}{A^2} \hat{\nabla}_{\alpha} \Pi_{\mu}^{\alpha}
% \end{equation}

We define the perturbation of energy density about the exact zero-order FRW metric by the comoving first order quantity $X_{\mu}^{(i)} \equiv \hat{\nabla}_{\mu} \rho^{(i)} \equiv \rho^{(i)}\Delta^{(i)}_\mu$ in the Einstein and $\tilde {X}_{\mu}^{(i)} \equiv \hat{\tilde \nabla}_{\mu} \tilde \rho^{(i)} \equiv \tilde \rho^{(i)}\tilde{\Delta}^{(i)}_\mu$ in the Jordan frame. These quantities are related by
\begin{eqnarray}\label{comoving spacial derivative of energy density}
\tilde {X}_{\mu}^{(i)} &=& \frac{1}{A^2}X_{\mu}^{(i)} - 2 \frac{\hat{\nabla}_{\mu} A}{A^3},\nonumber\\
\tilde {\Delta}_{\mu}^{(i)} &=& \Delta_{\mu}^{(i)} - 2 \frac{\hat{\nabla}_{\mu} A}{A},
\end{eqnarray}
and their time derivatives satisfy
\begin{widetext}
\begin{equation}\label{time derivative of comoving spacial derivative of energy density}
\mathring{\tilde{\Delta}}_{\mu}^{(i)} = \frac{1}{A^{\frac{1}{2}}} \dot{\Delta}_{\mu}^{(i)} - \frac{1}{2} \frac{\dot A}{A^{\frac{3}{2}}}\Delta_{\mu} - \frac{2}{A^{\frac{3}{2}}}\hat{\nabla}_{\mu}\dot{A} + \frac{2}{3A^{\frac{3}{2}}}\theta\hat{\nabla}_{\mu}A + 3\frac{\dot{A}}{A^{\frac{5}{2}}}\hat{\nabla}_{\mu}A - 2\frac{\dot A}{A^{\frac{3}{2}}}w_{\mu}.
\end{equation}
\end{widetext}

%*********** With X_µ = a/rho \nabla_µ \rho and \tilde X_µ = \tilde a/rho \tilde \nabla_µ \tilde rho

%\begin{equation}\label{comoving spacial derivative of energy density}
%\tilde {\mathcal X}_{\mu}^{(i)} = \sqrt A \mathcal X_{\mu}^{(i)} - 2 a \frac{\hat{\nabla}_{\mu} A}{\sqrt A}
%\end{equation}

%\begin{multline}\label{time derivative of comoving spacial derivative of energy density}
%\mathring{\tilde {\mathcal X}}_{\mu}^{(i)} = \dot {\mathcal X}_{\mu}^{(i)} - 2 a \frac{\hat{\nabla}_{\mu} \dot A}{A} + \frac{2}{3} a \theta \frac{\hat{\nabla}_{\mu} A}{A} - 2 a \dot u_{\mu} \frac{\dot A}{A} \\ + 2 a \frac{\dot A}{A^2} \hat{\nabla}_{\mu} A - 2 \dot a \frac{\hat{\nabla}_{\mu} A}{A}
%\end{multline}
%***********
Finally, it is useful to have the following expressions, which relate the conservations of the stress-energy tensors in the Einstein and the Jordan frames,
\begin{eqnarray}\label{divergence of energy momentum tensor}
\tilde \nabla_{\mu} \tilde T^{\mu \nu}_{(i)} &=& \frac{1}{A^3} \nabla_{\mu} T^{\mu \nu}_{(i)} - \frac{1}{2}\frac{\nabla^{\nu} A}{A^4}T_{(i)},\nonumber\\ 
\nabla_{\mu} \tilde T^{\mu (i)}_{\ \ \nu} &=& \frac{1}{A^2} \nabla_{\mu} T^{\mu (i)}_{\ \ \nu} - \frac{1}{2} \frac{\nabla_{\nu} A}{A^3} T^{(i)}.
\end{eqnarray}

\subsubsection{Einstein's equations}

We can now discuss the perturbed Einstein equations in the two frames. In $3+1$ formalism, the linearised Einstein equations give the relations between the kinematic and dynamical quantities that were introduced above. These include five constraint equations
\begin{eqnarray}\label{c1}
0 &=& \hat{\nabla}^{\alpha}\left( \epsilon^{\mu\nu}_{\ \ \ \alpha\beta}u^\beta\varpi_{\mu\nu}\right), \\
\label{c2}
0 &=& -\frac{2\hat{\nabla}_\mu\theta}{3} + \hat{\nabla}_\nu\sigma^{\nu}_{\ \mu} + \hat{\nabla}_\nu\varpi^{\nu}_{\ \mu} + M_{\rm Pl}^{-2}q_{\mu},\\
\label{c3}
0 &=& \left[ \hat{\nabla}^\alpha\sigma_{\beta(\mu} - \hat{\nabla}^\alpha\varpi_{\beta(\mu}\right]\epsilon_{\nu)\gamma\alpha}^{\ \ \ \ \beta} u^\gamma - H_{\mu\nu}, \\
\label{c4}
0 &=& \hat{\nabla}_\nu{E}^{\nu}_{\ \mu} + \frac{1}{2}M^{-2}_{\rm Pl}\left[ \hat{\nabla}_{\nu}\Pi^{\nu}_{\ \mu} + \frac{2}{3}\theta q_{\mu} - \frac{2}{3}\hat{\nabla}_\mu\rho\right], \\
\label{c5}
0 &=& \hat{\nabla}^\nu{H}_{\mu\nu} + \frac{1}{2}M^{-2}_{\rm Pl}\left[ \hat{\nabla}_\alpha q_\beta + (\rho + p)\varpi_{\alpha\beta}\right]\epsilon_{\mu\nu}^{\ \ \alpha\beta}u^\nu,\ \ \ \
\end{eqnarray}
and five propagation equations
\begin{widetext}
\begin{eqnarray}\label{p1}
0 &=& \dot{\theta} + \frac{1}{3}\theta^2 - \hat{\nabla} \cdot w + \frac{1}{2}M^{-2}_{\rm Pl}(\rho + 3p), \\
\label{p2}
0 &=& \dot{\sigma}^{\mu}_{\ \nu} + \frac{2}{3}\theta\sigma^{\mu}_{\ \nu} - \hat{\nabla}^{\langle\mu}w_{\nu\rangle} + {E}^{\mu}_{\ \nu} - \frac{1}{2}M^{-2}_{\rm Pl}\Pi^{\mu}_{\ \nu}, \\
\label{p3}
0 &=&\dot{\varpi}_{\mu\nu} + \frac{2}{3}\theta\varpi_{\mu\nu} - \hat{\nabla}_{[\mu}w_{\nu]}, \\
\label{p4}
0 &=& \dot{{E}}^{\mu}_{\ \nu} + \theta{E}^{\mu}_{\ \nu} - \hat{\nabla}^\alpha{H}_{\beta}^{\ (\mu}\epsilon_{\nu)\gamma\alpha}^{\ \ \ \ \ \ \ \beta}u^\gamma + \frac{1}{2}M^{-2}_{\rm Pl}\left[ \dot{\Pi}^{\mu}_{\ \nu} + \frac{1}{3}\theta\Pi^{\mu}_{\ \nu}\right] + \frac{1}{2}M^{-2}_{\rm Pl}\left[ (\rho + p)\sigma^{\mu}_{\ \nu} + \hat{\nabla}^{\langle\mu}q_{\nu\rangle}\right], \\
\label{p5}
0 &=&\dot{{H}}_{\mu\nu} + \theta{H}_{\mu\nu} + \hat{\nabla}^\alpha{E}_{\beta(\mu}\epsilon_{\nu)\gamma\alpha}^{\ \ \ \ \ \beta}u^\gamma  - \frac{1}{2}M^{-2}_{\rm Pl} \hat{\nabla}^\alpha\Pi_{\beta(\mu}\epsilon_{\nu)\gamma\alpha}^{\ \ \ \ \ \beta}u^\gamma.
\end{eqnarray}
\end{widetext}
In these equations, $\epsilon_{\mu\nu\alpha\beta}$ is the 4-dimensional covariant permutation tensor, $\hat{\nabla} \cdot w \equiv \hat{\nabla}^\alpha w_\alpha$ (the same for general vectors), and ${E}_{\mu\nu}$ and ${H}_{\mu\nu}$ are, respectively, the electric and magnetic parts of the Weyl tensor, ${C}_{\mu\nu\alpha\beta}$, defined by ${E}_{\mu\nu} \equiv u^\alpha u^\beta{C}_{\mu\alpha\nu\beta}$ and ${H}_{\mu\nu} \equiv \frac{1}{2}u^\alpha u^\beta \epsilon_{\mu\alpha}^{\ \ \gamma\delta}{C}_{\gamma\delta\nu\beta}$.

In the Jordan frame, these equations take the same form, but the quantities in them should become tildered. In addition, since there are extra terms in the Jordan-frame {\it effective} stress-energy tensor from the conformal transformation, as shown in Eq.~\eqref{eq:Tmn_eff}, such terms must be added to the tildered (bold) dynamical quantities in the Einstein equations. For simplicity, we do not repeat all the constraint and propagation equations, but instead only write down those that are directly relevant for linear perturbation evolutions in a spatially-flat perturbed universe: 
\begin{eqnarray}
\label{eq:c2-JF} 0 &=& -\frac{2\hat{\tilde{\nabla}}_\mu\tilde{\theta}}{3} + \hat{\tilde{\nabla}}_\nu\tilde{\sigma}^{\nu}_{\ \mu} + \hat{\tilde{\nabla}}_\nu\tilde{\varpi}^{\nu}_{\ \mu} + \tilde{M}_{\rm Pl}^{-2}\tilde{\bf q}_{\mu}, \\
\label{eq:c4-JF}
0 &=& \hat{\tilde{\nabla}}_\nu{\tilde{E}}^{\nu}_{\ \mu} + \frac{1}{2}\left[\hat{\tilde{\nabla}}_{\nu}\frac{\tilde{\bf \Pi}^{\nu}_{\ \mu}}{\tilde{M}^2_{\rm Pl}} + \frac{2}{3}\tilde{\theta}\frac{\tilde{\bf q}_{\mu}}{\tilde{M}_{\rm Pl}^2} - \frac{2}{3}\hat{\tilde{\nabla}}_\mu\frac{\tilde{\boldsymbol{\rho}}}{\tilde{M}^2_{\rm Pl}}\right], \\
\label{eq:p1-JF} 0 &=& \mathring{\tilde \theta} + \frac{1}{3}{\tilde \theta}^2 - \hat{\tilde \nabla} \cdot \tilde w + \frac{1}{2}{\tilde M}^{-2}_{\rm Pl}(\tilde{\boldsymbol{\rho}} + 3 \tilde{\bf p}), \\
\label{eq:p2-JF} 0 &=& \mathring{\tilde \sigma}^{\mu}_{\ \nu} + \frac{2}{3}\tilde\theta\tilde\sigma^{\mu}_{\ \nu} - \hat{\tilde \nabla}^{\langle\mu}\tilde w_{\nu\rangle} + {\tilde E}^{\mu}_{\ \nu} - \frac{1}{2}\tilde M^{-2}_{\rm Pl}\tilde{{\bf \Pi}}^{\mu}_{\ \nu},
\end{eqnarray}
and
\begin{widetext}
\begin{eqnarray}
\label{eq:p4-JF}
0 &=& \mathring{\tilde{E}}^{\mu}_{\ \nu} + \tilde{\theta}\tilde{E}^{\mu}_{\ \nu} - \hat{\tilde{\nabla}}^\alpha{\tilde{H}}_{\beta}^{\ (\mu}\epsilon_{\nu)\gamma\alpha}^{\ \ \ \ \ \beta}\tilde{u}^\gamma + \frac{1}{2}\left[ \left(\tilde{M}^{-2}_{\rm Pl}\tilde{\bf \Pi}^{\mu}_{\ \nu}\right)^{\circ} + \frac{1}{3}\tilde{\theta}\tilde{M}^{-2}_{\rm Pl}\tilde{\bf \Pi}^{\mu}_{\ \nu}\right] + \frac{1}{2}\left[ \tilde{M}^{-2}_{\rm Pl}(\tilde{\boldsymbol\rho} + \tilde{\bf p})\tilde{\sigma}^{\mu}_{\ \nu} + \hat{\tilde{\nabla}}^{\langle\mu}\tilde{M}^{-2}_{\rm Pl}\tilde{\bf q}_{\nu\rangle}\right].
\end{eqnarray}
\end{widetext}
We have verified that these Jordan frame equations are equivalent to their Einstein-frame counterparts (for example, one can start from the Jordan frame equations and use the relations of the dynamical and kinematic quantities in these two frames to derive the Einstein-frame equations, and vice versa), for which we have used the following expressions (up to linear order)
\begin{eqnarray}
\tilde{E}_{\mu\nu} &=& E_{\mu\nu},\\
\hat{\tilde{\Box}}A &=& \frac{1}{A}\hat{\Box}A,\\
\hat{\tilde{\nabla}}^{\nu}\hat{\tilde{\nabla}}_{\langle\mu}\hat{\tilde{\nabla}}_{\nu\rangle}A &=& \frac{2}{3A}\hat{\nabla}_{\mu}\hat{\Box}A - \frac{\dot{A}}{A}\hat{\nabla}_{\nu}\varpi^{\nu}_{\ \mu},\\
\hat{\tilde \nabla}_{\mu} \tilde w_{\nu} &=& \hat{\nabla}_{\mu} w_{\nu} + \frac{1}{2A} \hat{\nabla}_{\langle \mu} \hat{\nabla}_{\nu \rangle} A + \frac{1}{6A} h_{\mu \nu} \hat{\Box} A,~~~~
\end{eqnarray}
and
\begin{widetext}
\begin{eqnarray}
\hat{\tilde{\nabla}}_{\mu}\left[\frac{\tilde{\boldsymbol{\rho}}^{(\varphi)}}{\tilde{M}^{2}_{\rm Pl}}\right] &=& \frac{1}{A}\hat{\nabla}_{\mu}\left[\frac{\rho^{(\varphi)}}{M^{2}_{\rm Pl}}\right] - \frac{1}{A^2}\frac{\rho^{(\varphi)}}{M^{2}_{\rm Pl}}\hat{\nabla}_{\mu}A + \frac{\dot{A}}{A^2}Z_{\mu} + \frac{1}{A^2}\left[\theta+\frac{3}{2}\frac{\dot{A}}{A}\right]\hat{\nabla}_{\mu}\dot{A} - 2\theta\frac{\dot{A}}{A^3}\hat{\nabla}_{\mu}A - \frac{9}{4}\frac{\dot{A}^2}{A^4}\hat{\nabla}_{\mu}A - \frac{1}{A^2}\hat{\nabla}_{\mu}\hat{\Box}A.~~~~~~
\end{eqnarray}
\end{widetext}

In addition to the above equations, it is often useful to express the projected Ricci scalar, $\hat{R}$, onto the hypersurfaces orthogonal to $u^\mu$, as
\begin{eqnarray}\label{spatial_curvature0}
\hat{R} &=& R-2\dot{\theta}-\frac{4}{3}\theta^2+2\hat{\nabla}\cdot{w}\\
\label{spatial_curvature} &=& 2M^{-2}_{\rm Pl}\rho - \frac{2}{3}\theta^2.
\end{eqnarray}
The covariant spatial derivative of the projected Ricci scalar, $\eta_\mu \equiv \hat{\nabla}_\mu\hat{R}/2$, can be derived from the above equation, as
\begin{eqnarray}\label{derivative_spatial_curvature}
\eta_\mu &=& M^{-2}_{\rm Pl} \hat{\nabla}_\mu\rho - \frac{2}{3}\theta\hat{\nabla}_\mu\theta,
\end{eqnarray}
and its time evolution is governed by the following propagation equation
\begin{eqnarray}\label{propagation_spatial_curvature}
\dot{\eta}_\mu + \theta\eta_\mu &=& -\frac{2\theta}{3}\hat{\nabla}_\mu\hat{\nabla}\cdot w - M^{-2}_{\rm Pl}\hat{\nabla}_\mu\hat{\nabla}\cdot q.
\end{eqnarray}
% Baojiu: because of the definition of eta_mu, the coefficient in front of eta_mu becomes theta rather than 2/3*theta as in the standard case

In the Jordan frame, using the relations Eqs.~(\ref{spatial_curvature0}) and \eqref{eq:R-JF-EF}, it can be found that
\begin{eqnarray}
\hat{\tilde{R}} &=& \frac{1}{A}\hat{R} - \frac{2}{A^2}\hat{\Box}A,\\
\tilde{\eta}_{\mu} &=& \frac{1}{A}\eta_{\mu} - \frac{1}{A^2}\hat{\nabla}_{\mu}\hat{\Box}A,
\end{eqnarray}
with which it can be easily shown that the equations
\begin{eqnarray}
\hat{\tilde{R}} &=& 2\tilde{M}^2_{\rm Pl}\tilde{\boldsymbol{\rho}} - \frac{2}{3}\tilde{\theta}^2,\\
\label{derivative_spatial_curvature-JF}\tilde{\eta}_{\mu} &=& \tilde{M}^{-2}_{\rm Pl}\hat{\tilde{\nabla}}_{\mu}\tilde{\boldsymbol{\rho}} - \frac{2}{3}\tilde{\theta}\hat{\tilde{\nabla}}_{\mu}\tilde{\theta},\\
\label{propagation_spatial_curvature_JF}\dot{\tilde{\eta}}_\mu + \tilde{\theta}\tilde{\eta}_\mu &=& -\frac{2\tilde{\theta}}{3}\hat{\tilde{\nabla}}_\mu\hat{\tilde{\nabla}}\cdot\tilde{w} - \hat{\tilde{\nabla}}_\mu\left[\tilde{M}^{-2}_{\rm Pl}\hat{\tilde{\nabla}}\cdot\tilde{\bf q}\right].
\end{eqnarray}
are equivalent to their Einstein-frame counterparts.

\subsubsection{Conservation equations}

The Jordan frame stress-momentum tensor $\tilde{T}^{\mu}_{\ \nu}$ satisfies the conservation equation $\tilde{\nabla}_{\nu}\tilde{T}^{\nu}_{\ \mu}=0$, which can be decomposed into a component parallel to $\tilde{u}^{\mu}$ (the continuity equation) plus a component perpendicular (the Euler equation) as: 
\begin{eqnarray}\label{eq:continuity-JF}
\mathring{\tilde{\rho}} + (\tilde{\rho} + \tilde{P})\tilde{\theta} + \hat{\tilde{\nabla}}\cdot\tilde{q} &=& 0,\\
\label{eq:Euler-JF}
\mathring{\tilde{q}}_\mu + \frac{4}{3}\tilde{\theta}\tilde{q}_{\mu} + (\tilde{\rho}+\tilde{P})\tilde{w}_\mu + \hat{\tilde{\nabla}}_\mu\tilde{P} + \hat{\tilde{\nabla}}_\nu\tilde{\Pi}^{\nu}_{\ \mu} &=& 0.
\end{eqnarray}

In the Einstein frame, on the other hand, the stress-energy tensors for individual species do not conserve, but satisfy 
\begin{eqnarray}
\nabla_{\nu}T^{\nu}_{\ \mu} &=& \frac{T}{2A}\nabla_{\mu}A,\nonumber
\end{eqnarray}
according to Eq.~\eqref{divergence of energy momentum tensor}, where $T$ is the trace of $T^{\mu}_{\ \nu}$. Therefore, the continuity and Euler equations can be written as
\begin{eqnarray}\label{eq:continuity-EF}
\dot{{\rho}} + ({\rho} + {P}){\theta} + \hat{{\nabla}}\cdot{q} &=& -\frac{T}{2A}\dot{A},\\
\label{eq:Euler-EF}
\dot{{q}}_\mu + \frac{4}{3}{\theta}{q}_{\mu} + ({\rho}+{P}){w}_\mu + \hat{{\nabla}}_\mu{P} + \hat{{\nabla}}_\nu{\Pi}^{\nu}_{\ \mu} &=& \frac{T}{2A}\hat{\nabla}_{\mu}A.~~~~~~~~
\end{eqnarray}
In what follows we shall explicitly compare the conservation equation for photons and dark matter in the two frames. Other matter species, e.g., massless neutrinos and baryons, are similar. We will not discuss massive neutrinos in this paper.

Photons (and massless particles in general) have zero trace of their stress-energy tensor, and so the conservation equations hold for them even in the Einstein frame. By using Eqs.~\eqref{expansion scalar}, \eqref{energy density}, \eqref{time derivative of energy density} and $\hat{\tilde{\nabla}}\cdot\tilde{q}=A^{-5/2}\hat{\nabla}\cdot{q}$, one can straightforwardly check that the continuity equations for photons,
\begin{eqnarray}\label{energy conservation equation jf}
\mathring{\tilde \rho}^{(\gamma)} + \frac{4}{3} \tilde \theta \tilde \rho^{(\gamma)} + \hat{\tilde \nabla}_{\alpha} \tilde q^{\alpha}_{(\gamma)} &=& 0,\\
\label{energy conservation equation ef}
\dot \rho^{(\gamma)} + \frac{4}{3} \theta \rho^{(\gamma)} + \hat{\nabla}_{\alpha} q^{\alpha}_{(\gamma)} &=& 0,
\end{eqnarray}
are equivalent to each other in the two frames.
% has the same form as the one expressed in the Einstein frame, with all tildered quantities replaced by their untildered counterparts :
% \begin{equation}\label{energy conservation equation jf}
% \dot \rho^{(\gamma)} + \frac{4}{3} \theta \rho^{(\gamma)} + \hat{\nabla}_{\alpha} q^{\alpha}_{(\gamma)} = 0
% \end{equation}
% Using \eqref{time derivative of energy flux}, \eqref{time derivative of expansion scalar}, \eqref{energy flux}, \eqref{spatial derivative of anisotropic stress}, \eqref{energy density}, \eqref{acceleration}, \eqref{spatial derivative of energy density}, \eqref{Thomson cross section} and \eqref{velocity}, one can check that the momentum conservation equation (geodesic equation) for photons in the Jordan frame :
Similarly, it can be checked that the momentum (Euler) equations for photons in the two frames,
\begin{widetext}
\begin{eqnarray}
\mathring{\tilde q}_{\mu}^{(\gamma)} + \frac{4}{3} \tilde \theta \tilde q_{\mu}^{(\gamma)} + \hat{\tilde \nabla}_{\nu} \tilde \Pi_{\mu}^{\nu(\gamma)} + \frac{4}{3} \tilde \rho^{(\gamma)} {\tilde w}_{\mu} + \frac{1}{3} \hat{\tilde \nabla}_{\mu} \tilde \rho^{(\gamma)} &=& \tilde n_e \tilde \sigma_T \left( \frac{4}{3} \tilde \rho^{(\gamma)} \tilde v_{\mu}^{(b)} - \tilde q_{\mu}^{(\gamma)} \right),\\
\dot q_{\mu}^{(\gamma)} + \frac{4}{3} \theta q_{\mu}^{(\gamma)} + \hat{\nabla}_{\nu} \Pi_{\mu}^{\nu(\gamma)} + \frac{4}{3} \rho^{(\gamma)} w_{\mu} + \frac{1}{3} \hat{\nabla}_{\mu} \rho^{(\gamma)} &=& n_e \sigma_T \left( \frac{4}{3} \rho^{(\gamma)} v_{\mu}^{(b)} - q_{\mu}^{(\gamma)} \right),
\end{eqnarray}
\end{widetext}
are also equivalent, for which we have used
\begin{eqnarray}
\mathring{\tilde{q}}_{\mu} &=& \frac{1}{A^2}\dot{q}_{\mu} - 2\frac{\dot{A}}{A^3}q_{\mu},\nonumber\\
\tilde{v}_{\mu} &=& A^{\frac{1}{2}}v_{\mu},\nonumber\\
\tilde{n}_e\tilde{\sigma}_{\rm T} &=& A^{-\frac{1}{2}}n_e\sigma_{\rm T},\nonumber
\end{eqnarray}
where $n_e$ is the electron number density and $\sigma_{\rm T}$ is the Thomson scattering cross section. The right-hand sides of the Euler equations are the interactions between electrons and photons. Note that the electron densities in the two frames are related by $\tilde{n}_e = A^{-\frac{3}{2}}n_e$, because the electron numbers are the same while the volumes are related by $\tilde{V}\propto\tilde{l}^3=(\sqrt{A}l)^3\propto V$. On the other hand, the Thomson cross-sections in the two frames are connected by $\tilde \sigma_T = A \ \sigma_T$, which can be easily checked using the expression $\sigma_T = \frac{8\pi}{3} \left(\frac{q^2}{4\pi \epsilon_0 m c^2} \right)^2$, in which $q$ and $m$ are respectively the electron electric charge and the mass; thus $\sigma_{\rm T}\propto m^{-2}$ with $\tilde{m}=A^{-\frac{1}{2}}m$.

% has the same form as the one expressed in the Einstein frame, with all tildered quantities replaced by their untildered counterparts : 

% \begin{multline}
% \dot q_{\mu}^{(\gamma)} + \frac{4}{3} \theta q_{\mu}^{(\gamma)} + \hat{\nabla}_{\alpha} \Pi_{\mu}^{\alpha(\gamma)} + \frac{4}{3} \rho^{(\gamma)} \dot u_{\mu} + \frac{1}{3} \hat{\nabla}_{\mu} \rho^{(\gamma)} \\ = n_e \sigma_T \left( \frac{4}{3} \rho^{(\gamma)} v_{\mu}^{(b)} - q_{\mu}^{(\gamma)} \right)
% \end{multline}

By taking the spatial gradients of Eqs.~\eqref{energy conservation equation jf} and \eqref{energy conservation equation ef}, one finds the propagation equations for the photon density contrast $\tilde{\Delta}_{\mu}$ ($\Delta_{\mu}$) in the Jordan (Einstein) frame (using Eq.~\eqref{time derivative of comoving spacial derivative of energy density})
\begin{eqnarray}\label{propagation equation for the fractional spatial gradient of the photon density in the Einstein frame}
\label{eq:gamma_delta_JF}\mathring{\tilde{\Delta}}_{\mu}^{(\gamma)} + \frac{1}{3}\tilde{\theta}\tilde{\Delta}_{\mu} + \frac{4}{3}\tilde {{Z}}_{\mu} + \frac{4}{3} \tilde{\theta}\tilde{w}_{\mu} + \frac{1}{\tilde \rho^{(\gamma)}} \hat{\tilde \nabla}_{\mu} \hat{\tilde \nabla}_{\alpha} \tilde q^{\alpha}_{(\gamma)} &=& 0,\\
\label{eq:gamma_delta_EF}\dot{\Delta}_{\mu}^{(\gamma)} + \frac{1}{3}{\theta}{\Delta}_{\mu} + \frac{4}{3} {Z}_{\mu}  + \frac{4}{3}{\theta}{w}_{\mu} + \frac{1}{\rho^{(\gamma)}} \hat{\nabla}_{\mu} \hat{\nabla}_{\alpha} q^{\alpha}_{(\gamma)} &=& 0.~~~~~~~~
% \mathring{\tilde{\Delta}}_{\mu}^{(\gamma)} + \frac{4}{3} \tilde {{Z}}_{\mu} + \frac{\tilde a}{\tilde \rho^{(\gamma)}} \hat{\tilde \nabla}_{\mu} \hat{\tilde \nabla}_{\alpha} \tilde q^{\alpha}_{(\gamma)} + \frac{4}{3} \tilde a \tilde{\theta}\tilde{w}_{\mu} &=& 0,\\
% \dot{\Delta}_{\mu}^{(\gamma)} + \frac{4}{3} {Z}_{\mu} + \frac{a}{\rho^{(\gamma)}} \hat{\nabla}_{\mu} \hat{\nabla}_{\alpha} q^{\alpha}_{(\gamma)} + \frac{4}{3} a{\theta}{w}_{\mu} &=& 0.
\end{eqnarray}
Again, it can be checked that these equations in the two frames are equivalent. The fact that all the conservation equations for photons take exactly the same form in the Jordan and Einstein frames is because conformal coupling does not affect the dynamics of photons and, more generally, massless particles.
% Using eqs. \eqref{time derivative of comoving spacial derivative of energy density}, \eqref{comoving spatial derivative of expansion scalar}, \eqref{energy density} and \eqref{spatial derivative of spatial derivative of energy flux}, one can check that eq \eqref{propagation equation for the fractional spatial gradient of the photon density in the Einstein frame} gives the propagation equation for the comoving fractional spatial gradient of the photon density $\mathcal{X}_{\mu}$ in the Einstein frame :

% \begin{equation}
% \dot{\mathcal{X}}_{\mu}^{(\gamma)} + \frac{4}{3} \mathcal{Z}_{\mu} + \frac{a}{\rho^{(\gamma)}} \hat{\nabla}_{\mu} \hat{\nabla}_{\alpha} q^{\alpha}_{(\gamma)} + \frac{4}{3} a \theta \dot u_{\mu} = 0
% \end{equation}

% This latter equation can be directly obtained by taking the spatial gradient $\hat{\nabla}_{\mu}$ of eq. \eqref{energy conservation equation jf}.\\

We next turn to cold dark matter, which is treated as a perfect fluid with zero pressure and anisotropic stress in the perturbation analysis. Using the connection between quantities in the two frames, it can be shown that the continuity equations in the Jordan and Einstein frames:
\begin{eqnarray}\label{energy conservation equation-cdm jf}
\mathring{\tilde \rho}^{(\rm c)} + \tilde \theta \tilde \rho^{(\rm c)} + \hat{\tilde \nabla}_{\alpha} \tilde q^{\alpha}_{(\rm c)} &=& 0,\\
\label{energy conservation equation-cdm ef}
\dot \rho^{(\rm c)} + \left[\theta-\frac{\dot{A}}{2A}\right]\rho^{(\rm c)} + \hat{\nabla}_{\alpha} q^{\alpha}_{(\rm c)} &=& 0,
\end{eqnarray}
are equivalent to each other. Note that, unlike the case of photons, these equations take slightly different forms in the Jordan and Einstein frames, with the latter having an additional term. While in the Jordan frame $\tilde{\rho}^{(\rm c)}$ satisfies the usual $\tilde{\rho}^{(\rm c)}\propto\tilde{a}^{-3}$ scaling law, in the Einstein frame the mass of dark matter particles evolves in time with $m\propto\sqrt{A}$, and we have a modified scaling law $\rho^{(\rm c)}/\sqrt{A}\propto{a}^{-3}$ or $\rho^{(\rm c)}\propto\sqrt{A}a^{-3}$, which explains the extra factor in Eq.~\eqref{energy conservation equation-cdm ef}.

% In the Jordan frame, projecting the equation $\tilde \nabla_{\mu} \tilde T^{\mu}_{\nu} = 0$ along the observer's 4-velocity $\tilde u^{\mu}$ gives the energy conservation equation (continuity equation) for CDM :

% \begin{equation}\label{CDM continuity eq JF}
% \mathring{\tilde \rho} + \tilde \theta \tilde \rho + \hat{\tilde \nabla}_{\alpha} \tilde q^{\alpha} = 0	
% \end{equation}

% In the Einstein frame, the divergence of the stress energy tensor is no longer equal to zero, but given by eq. \eqref{divergence of energy momentum tensor}, where $\tilde \nabla_{\mu} \tilde T^{\mu}_{\nu} = 0$. Let's project it along $u^{\mu}$ to get the energy conservation equation :

% \begin{equation}\label{CDM continuity eq EF}
% \dot \rho + \theta \rho + \hat{\nabla}_{\alpha} q^{\alpha} = \pm \frac{\rho}{2} \frac{\dot A}{A}
% \end{equation}

Similarly, the momentum equations in the two frames:
\begin{eqnarray}\label{Euler eq-cdm JF}
\mathring{\tilde{q}}^{(\rm c)}_{\mu} + \frac{4}{3}\tilde{\theta}\tilde{q}^{(\rm c)}_{\mu} + \tilde{\rho}^{(\rm c)}\tilde{w}_{\mu} &=& 0,\nonumber\\
\label{Euler eq-cdm EF}
\dot{q}^{(\rm c)}_{\mu} + \frac{4}{3}\theta{q}^{(\rm c)}_{\mu} + \rho^{(\rm c)}{w}_{\mu} &=& -\frac{1}{2A} \rho^{(\rm c)}\hat{\nabla}_{\mu} A,\nonumber
\end{eqnarray}
are equivalent to each other. Indeed, the above equations can be rewritten in terms of the peculiar velocities given by $\tilde{v}^{{\rm c}}_{\mu}\equiv\tilde{q}^{(\rm c)}_{\mu}/\tilde{\rho}^{(\rm c)}$ and $v^{(\rm c)}_{\mu}\equiv{q}^{(\rm c)}_{\mu}/\rho^{(\rm c)}$:
\begin{eqnarray}\label{veq-cdm JF}
\mathring{\tilde{v}}^{(\rm c)}_{\mu} + \frac{\mathring{\tilde{a}}}{\tilde{a}}\tilde{v}^{(\rm c)}_{\mu} + \tilde{w}_{\mu} &=& 0,\\
\label{veq-cdm EF}\dot{v}^{(\rm c)}_{\mu} + \left[\frac{\dot{a}}{a}+\frac{\dot{A}}{2A}\right]{v}^{(\rm c)}_{\mu} + {w}_{\mu} &=& -\frac{1}{2A}\hat{\nabla}_{\mu}A.
\end{eqnarray}
As expected, in the Jordan frame, the peculiar velocity of dark matter particles is affected by two terms -- a `frictional' force caused by the expansion of the Universe (the second, velocity-dependent term in Eq.~\eqref{veq-cdm JF}), and a 4-acceleration $\tilde{w}_{\mu}$ which encodes the effect of gravity on particle geodesics. The terms are both modified in the Einstein frame: the particles now feel a `fifth' force that is proportional to the gradient\footnote{In this sense $\ln(A)$ can be considered as the potential of the fifth force.} of $\ln(A)$ as in the right-hand side of Eq.~\eqref{veq-cdm EF}, and an additional frictional force as in the brackets on the left side of Eq.~\eqref{veq-cdm EF}.

Taking the spatial gradients of Eqs.~\eqref{energy conservation equation-cdm jf} and \eqref{energy conservation equation-cdm ef}, we obtain the propagation equations for the dark matter density contrasts in the two frames (again by using Eq.~\eqref{time derivative of comoving spacial derivative of energy density}):
\begin{eqnarray}\label{propagation equation for the fractional spatial gradient of the cdm density in the Jordan frame}
\mathring{\tilde{\Delta}}_{\mu}^{(\rm c)} + \frac{1}{3}\tilde{\theta}\tilde{\Delta}^{(\rm c)}_{\mu} + \tilde{{Z}}_{\mu} + \tilde{\theta}\tilde{w}_{\mu} + \frac{1}{\tilde \rho^{(\rm c)}} \hat{\tilde \nabla}_{\mu} \hat{\tilde \nabla}_{\alpha} \tilde q^{\alpha}_{(\rm c)} &=& 0,\\
\label{propagation equation for the fractional spatial gradient of the cdm density in the Einstein frame}\dot{\Delta}_{\mu}^{(\rm c)} + \frac{1}{3}{\theta}{\Delta}^{(\rm c)}_{\mu} + {Z}_{\mu}  + {\theta}{w}_{\mu} + \frac{1}{\rho^{(\rm c)}} \hat{\nabla}_{\mu} \hat{\nabla}_{\alpha} q^{\alpha}_{(\rm c)} &=& \frac{1}{2A}\hat{\nabla}_{\mu}\dot{A} - \frac{\dot{A}}{2A^2}\hat{\nabla}_{\mu}A + \frac{\dot{A}}{2A}w_{\mu}.
\end{eqnarray}

Before leaving this section, note that another sanity check of the equations derived above is to verify that the components of the scalar field effective stress-energy tensor: $\tilde{\boldsymbol{\rho}}$, $\tilde{\bf p}$, $\tilde{\bf q}_{\mu}$ and $\tilde{\bf \Pi}^{\mu}_{\ \nu}$ satisfy the conservation equations, Eq.~\eqref{eq:continuity-JF} and \eqref{eq:Euler-JF}. An explicit check of this will require us to know the exact form of $\tilde{\rho}^{(\tilde \phi)}, \tilde{p}^{(\tilde \phi)}, \tilde{q}_{\mu}^{(\tilde \phi)}$ and $\tilde{\Pi}^{(\tilde \phi)}_{\mu\nu}$, as well the stress-energy tensor components for normal matter species (because it is the components of $\tilde{M}^{-2}_{\rm Pl}T_{\mu\nu}^{(\rm matter)}$ that enter the Jordan-frame Einstein equations, and $\tilde{M}_{\rm Pl}$ depends on the scalar field $A$). A slightly easier check -- which still serves the purpose -- is to assume that $\tilde{T}^{(\tilde \phi)}_{\mu\nu}=\tilde{T}_{\mu\nu}^{(\rm matter)}=0$ which means that there is no matter, including scalar field, in the Universe in the Einstein frame. We have checked that Eqs.~\eqref{eq:continuity-JF} and \eqref{eq:Euler-JF} hold for $\tilde{\boldsymbol{\rho}}$, $\tilde{\bf p}$, $\tilde{\bf q}_{\mu}$ and $\tilde{\bf \Pi}^{\mu}_{\ \nu}$ in this case (their Einstein-frame version are simply $0=0$ and so hold too trivially).

\subsection{Quantities and equations in $k$ space}

The linearised Einstein and conservation equations are usually solved in Fourier (or $k$) space, where the different Fourier (or $k$) modes are independent of each other. The spatial derivatives can then be replaced with multiplications by powers of $k$, so that the equations become ordinary differential equations that can be solved by numerical integration. In this subsection, we'll write the quantities and equations in $k$ space.

For this, we define the zero-order eigenfunctions $Q^{(k)}$ of the comoving spatial d'Alembertian operator $a^2\hat{\Box}\equiv{a}^2\hat{\nabla}_{\alpha} \hat{\nabla}^{\alpha}$ (and $\tilde{a}^2\hat{\tilde{\Box}}\equiv\tilde{a}^2\hat{\tilde{\nabla}}_{\alpha}\hat{\tilde{\nabla}}^{\alpha}$) as
\begin{equation}
a^2\hat{\Box} Q^{(k)} = -k^2 Q^{(k)}; \ \ \ \tilde{a}^2\hat{\tilde{\Box}}Q^{(k)} = -k^2Q^{(k)}.
\end{equation}
$\dot{Q}^{(k)}$ is a zero-order quantity. The multiplication of $a^2$ not only makes this operator a comoving one, but also means that $Q^{(k)}$ is the same for both the Einstein and the Jordan frames. Vector and (rank-2) tensor perturbation quantities in the Einstein (Jordan) frames can be expanded in terms of $Q_{\mu}^{(k)}$ ($\tilde{Q}_{\mu}^{(k)}$) and $Q_{\mu \nu}^{(k)}$ ($\tilde{Q}_{\mu\nu}^{(k)}$), which are defined by $Q_{\mu}^{(k)}\equiv\frac{a}{k}\hat{\nabla}_{\mu}Q^{(k)}$ ($\tilde{Q}_{\mu}^{(k)}\equiv\frac{\tilde{a}}{k}\hat{\tilde{\nabla}}_{\mu}Q^{(k)}$) and $Q_{\mu\nu}^{(k)}\equiv\frac{a}{k}\hat{\nabla}_{\langle\mu}Q_{\nu\rangle}^{(k)}$ ($\tilde{Q}_{\mu\nu}^{(k)}\equiv\frac{\tilde{a}}{k}\hat{\tilde{\nabla}}_{\langle\mu}\tilde{Q}_{\nu\rangle}^{(k)}$) respectively.

\subsubsection{Kinematic quantities}

Using the notations introduced above, the $k$-space kinematic quantities (or their gradients) in the two frames can be written as
\begin{widetext}
\begin{eqnarray}
&& Z_{\mu} = \sum_{k}\frac{k^2}{a^2}Z_{k}Q^{(k)}_{\mu},\ \ w_{\mu} = -\sum_{k}\frac{k}{a}w_{k}Q^{(k)}_{\mu},\ \ \sigma_{\mu\nu} = -\sum_{k}\frac{k}{a}\sigma_{k}Q^{(k)}_{\mu\nu}, \ \ 
h_{\mu} = \sum_{k}kh_{k}Q^{(k)}_{\mu},\ \ \eta_{\mu} = \sum_{k}\frac{k^3}{a^3}\eta_{k}Q^{(k)}_{\mu},\nonumber\\
&& \tilde{Z}_{\mu} = \sum_{k}\frac{k^2}{\tilde{a}^2}\tilde{Z}_{k}\tilde{Q}^{(k)}_{\mu},\ \ \tilde{w}_{\mu} = -\sum_{k}\frac{k}{\tilde{a}}\tilde{w}_{k}\tilde{Q}^{(k)}_{\mu},\ \ \tilde{\sigma}_{\mu\nu} = -\sum_{k}\frac{k}{\tilde{a}}\tilde{\sigma}_{k}\tilde{Q}^{(k)}_{\mu\nu},\ \ \tilde{h}_{\mu} = \sum_{k}k\tilde{h}_{k}\tilde{Q}^{(k)}_{\mu},\ \ \tilde{\eta}_{\mu} = \sum_{k}\frac{k^3}{\tilde{a}^3}\tilde{\eta}_{k}\tilde{Q}^{(k)}_{\mu},
\end{eqnarray}
\end{widetext}
where $h_{\mu}\equiv\hat{\nabla}_{\mu}a$ and $\tilde{h}_{\mu}\equiv\hat{\tilde{\nabla}}_{\mu}\tilde{a}$. From these relations we find
\begin{eqnarray}
\label{eq:wkexp}\tilde{w}_{k} &=& w_{k} - \frac{1}{2A}\xi_{k},\\
\label{eq:zkexp} k\tilde{Z}_{k} &=& kZ_{k} + \frac{3}{2}\frac{1}{A}\xi'_k - \frac{3}{2}\frac{1}{A}\left[\frac{a'}{a}+\frac{3A'}{2A}\right]\xi_k + \frac{3}{2}\frac{A'}{A}w_k,\\
\label{eq:sigmakexp}\tilde{\sigma}_{k} &=& \sigma_{k},\\
\tilde{h}_{k} &=& h_{k}+\frac{1}{2A}\xi_{k},\\
\label{eq:etakexp}\tilde{\eta}_{k} &=& \eta_{k} + \frac{1}{A}\xi_{k},
\end{eqnarray}
in which $'$ means the derivative with respect to the conformal time $\eta$ (not to be confused with $\eta_{k}$ with a subscript $_{k}$, which is the Fourier coefficient of $\eta_{\mu}$), and where we have used the Fourier expansion of $\hat{\nabla}_{\mu}A$ given as
\begin{eqnarray}
\hat{\nabla}_{\mu}A &=& \sum_{k}\frac{k}{a}\xi_{k}Q^{(k)}_{\mu},
\end{eqnarray}
and the relations (which hold to zero order)
\begin{eqnarray}\label{eq:QQ}
\frac{1}{\tilde{a}}\tilde{Q}^{(k)}_{\mu} = \frac{1}{a}Q^{(k)}_{\mu}, \ \ \frac{1}{\tilde{a}^2}\tilde{Q}^{(k)}_{\mu\nu} = \frac{1}{a^2}Q^{(k)}_{\mu\nu}.
\end{eqnarray}

\subsubsection{Dynamical quantities}

Similarly, for the dynamical quantities or their gradients, we can write
\begin{widetext}
\begin{eqnarray}
&& {\Delta}_{\mu} = \sum_{k}\frac{k}{{a}}\Delta_{k}{Q}^{(k)}_{\mu},\ \ q_{\mu} = -\sum_{k}q_{k}Q^{(k)}_{\mu},\ \ \Pi_{\mu\nu} = \sum_{k}\Pi_{k}Q^{(k)}_{\mu\nu},\nonumber\\
&& \tilde{\Delta}_{\mu} = \sum_{k}\frac{k}{\tilde{a}}\tilde{\Delta}_{k}\tilde{Q}^{(k)}_{\mu},\ \ \tilde{q}_{\mu} = -\sum_{k}\tilde{q}_{k}\tilde{Q}^{(k)}_{\mu},\ \ \tilde{\Pi}_{\mu\nu} = \sum_{k}\tilde{\Pi}_{k}\tilde{Q}^{(k)}_{\mu\nu}.
\end{eqnarray}
\end{widetext}
From these expressions we can find the following relations between the two frames
\begin{eqnarray}
\label{eq:Dk-transform}\tilde{\Delta}_{k} &=& \Delta_{k} - \frac{2}{A}\xi_{k},\\
\tilde{\Delta}^{p}_{k} &=& \Delta^{p}_{k} - \frac{2}{A}\xi_{k},\\
\tilde{q}_{k} &=& A^{-2}q_{k},\\
\tilde{\Pi}_{k} &=& A^{-2}\Pi_{k},
\end{eqnarray}
where $\Delta^{p}_{k}$ is the expansion coefficient for $\Delta_{\mu}^{p}\equiv\hat{\nabla}_{\mu}p/\rho$. Note that because $\tilde{\rho}=A^{-2}\rho$, if we define $v_{k}=q_{k}/\rho$ and $\pi_{k}=\Pi_{k}/\rho$ (and similarly for their Jordan-frame counterparts), we will have
\begin{eqnarray}
\tilde{v}_{k} = v_{k},\ \ \tilde{\pi}_{k} = \pi_{k}.
\end{eqnarray}

For the relevant effective stress-energy tensor components of the scalar field given in Eqs.~\eqref{scalar field energy density} to \eqref{scalar field anisotropic stress}, the Fourier expansion coefficient can be written as
\begin{eqnarray}
\label{eq:Xkexp}\frac{\tilde{\bf X}_{k}\tilde{a}^2}{\tilde{M}^{2}_{\rm Pl}} &=& \frac{X^{(\varphi)}_{k}a^2}{M^{2}_{\rm Pl}} - \frac{1}{A}\frac{\rho{a}^2}{M^{2}_{\rm Pl}}\xi_{k} + \frac{k^2}{A}\xi_{k} + \frac{A'}{A}kZ_{k} + \frac{3}{A}\left[\frac{a'}{a}+\frac{A'}{2A}\right]\xi'_{k} - \frac{6A'}{A^2}\left[\frac{a'}{a}+\frac{3}{8}\frac{A'}{A}\right]\xi_{k} + \frac{3A'}{A}\left[\frac{a'}{a}+\frac{A'}{2A}\right]w_{k},~~~\\
\label{eq:qkexp}\frac{\tilde{\bf q}_{k}\tilde{a}^2}{\tilde{M}^{2}_{\rm Pl}} &=& \frac{q^{(\varphi)}_{k}a^2}{M^{2}_{\rm Pl}} - \frac{1}{A}k\xi'_{k} + \frac{1}{A}\left[\frac{a'}{a}+\frac{3}{2}\frac{A'}{A}\right]k\xi_{k} - \frac{A'}{A}kw_{k},\\
\label{eq:pikexp}\frac{\tilde{\bf \Pi}_{k}\tilde{a}^2}{\tilde{M}^{2}_{\rm Pl}} &=& \frac{\Pi^{(\varphi)}_{k}a^2}{M^2_{\rm Pl}} - \frac{1}{A}k^2\xi_{k} - \frac{A'}{A}k\sigma_{k},
\end{eqnarray}
where in Eq.~\eqref{eq:Xkexp} $\rho$ is the total energy density of all matter species in the Einstein frame.

\subsubsection{Einstein's equations}

With the above results, we can now write down the Fourier-space versions of the linearised Einstein equations in the Jordan and Einstein frames, and check their equivalence. For the constraint equation, Eqs.~\eqref{eq:c2-JF} and \eqref{c2}, we have
\begin{eqnarray}\label{eq:c2-k}
\frac{2}{3}k^2\left(\tilde{\sigma}_{k}-\tilde{Z}_{k}\right) &=& \tilde{M}^{-2}_{\rm Pl}\tilde{\bf q}_{k}\tilde{a}^2,\nonumber\\
\frac{2}{3}k^2\left({\sigma}_{k}-{Z}_{k}\right) &=& {M}^{-2}_{\rm Pl}{q_{k}}{a}^2,
\end{eqnarray}
and using Eqs.~\eqref{eq:zkexp}, \eqref{eq:sigmakexp} and \eqref{eq:qkexp} it can be shown that they are equivalent. Using Eqs.~\eqref{eq:wkexp}, \eqref{eq:zkexp}, \eqref{eq:etakexp}, \eqref{eq:Xkexp} and \eqref{eq:qkexp}, it can also be found that the Fourier-space versions of Eqs.~\eqref{derivative_spatial_curvature-JF} and \eqref{derivative_spatial_curvature}:
\begin{eqnarray}
k^2\tilde{\eta}_{k} &=& \tilde{M}^{-2}_{\rm Pl}\tilde{\bf X}_{k}\tilde{a}^2 - 2k\frac{\tilde{a}'}{\tilde{a}}\tilde{Z}_{k},\nonumber\\
k^2{\eta}_{k} &=& {M}^{-2}_{\rm Pl}{X}_{k}{a}^2 - 2k\frac{{a}'}{a}{Z}_{k},
\end{eqnarray}
and of Eqs.~\eqref{propagation_spatial_curvature_JF} and \eqref{propagation_spatial_curvature}:
\begin{eqnarray}
k\tilde{\eta}'_{k} &=& -\tilde{M}^{-2}_{\rm Pl}\tilde{\bf q}_{k}\tilde{a}^2 - 2k\frac{\tilde{a}'}{\tilde{a}}\tilde{w}_{k},\nonumber\\
k{\eta}'_{k} &=& -{M}^{-2}_{\rm Pl}{q}_{k}{a}^2 - 2k\frac{{a}'}{{a}}{w}_{k},
\end{eqnarray}
are equivalent. 

The constraint equations Eqs.~\eqref{eq:c4-JF} and \eqref{c4} in the Fourier space become
\begin{eqnarray}
k^3\tilde{\Phi}_{k} &=& -\frac{1}{2}\left[k\tilde{M}^{-2}_{\rm Pl}\left(\tilde{\bf \Pi}_{k}+\tilde{\bf X}_{k}\right)+3\frac{\tilde{a}'}{\tilde{a}}\tilde{M}^{-2}_{\rm Pl}{\bf q}_{k}\right]\tilde{a}^2,\nonumber\\
k^3{\Phi}_{k} &=& -\frac{1}{2}\left[k{M}^{-2}_{\rm Pl}\left({\Pi}_{k}+{X}_{k}\right)+3\frac{a'}{a}{M}^{-2}_{\rm Pl}{q}_{k}\right]{a}^2,
\end{eqnarray}
the equivalence of which can be checked by using Eqs.~\eqref{eq:Xkexp}, \eqref{eq:qkexp}, \eqref{eq:pikexp} and \eqref{eq:c2-k}. In the above, $\Phi_{k}$ and $\tilde{\Phi}_{k}$ are respectively the Fourier coefficients of $E_{\mu\nu}$ and $\tilde{E}_{\mu\nu}$:
\begin{eqnarray}
E_{\mu\nu} = \sum_{k}\frac{k^2}{a^2}\Phi_{k}Q_{\mu\nu}, \ \ \tilde{E}_{\mu\nu} = \sum_{k}\frac{k^2}{\tilde{a}^2}\tilde{\Phi}_{k}\tilde{Q}_{\mu\nu},
\end{eqnarray}
so that the Weyl potentials in the two frames satisfy $\tilde{\Phi}_{k}=\Phi_{k}$ following $\tilde{E}_{\mu\nu}=E_{\mu\nu}$.

The propagation equations for the shear, Eqs.~\eqref{eq:p2-JF} and \eqref{p2}, become
\begin{eqnarray}
k\left(\tilde{\sigma}'_{k}+\frac{\tilde{a}'}{\tilde{a}}\tilde{\sigma}_{k}\right) - k^2\left(\tilde{\Phi}_{k}+\tilde{w}_{k}\right) + \frac{1}{2}\tilde{M}^{-2}_{\rm Pl}\tilde{\bf \Pi}_{k}\tilde{a}^2 &=& 0,\nonumber\\
k\left(\sigma'_{k}+\frac{a'}{a}\sigma_{k}\right) - k^2\left(\Phi_{k}+w_{k}\right) + \frac{1}{2}M^{-2}_{\rm Pl}\Pi_{k}{a}^2 &=& 0,
\end{eqnarray}
and the propagation equations for the Weyl potential, Eqs.~\eqref{eq:p4-JF} and \eqref{p4}, can be written in $k$-space as:
\begin{eqnarray}\label{eq:weyl_in_k}
k^2\left(\tilde{\Phi}'_{k}+\frac{\tilde{a}'}{\tilde{a}}\tilde{\Phi}_{k}\right) &=& \frac{1}{2}\left[k\tilde{M}^{-2}_{\rm Pl}\left[(\tilde{\boldsymbol{\rho}}+\tilde{\bf P})\tilde{\sigma}_{k}+\tilde{\bf q}_{k}\right]-\left(\tilde{M}^{-2}_{\rm Pl}\tilde{\bf \Pi}_{k}\right)' -\frac{\tilde{a}'}{\tilde{a}}\tilde{M}^{-2}_{\rm Pl}\tilde{\bf \Pi}_{k}\right]\tilde{a}^2,\nonumber\\
k^2\left(\Phi'_{k}+\frac{a'}{a}\Phi_{k}\right) &=& \frac{1}{2}\left[kM^{-2}_{\rm Pl}\left[(\rho+P)\sigma_{k}+q\right]-M^{-2}_{\rm Pl}\Pi'_{k}-\frac{a'}{a}M^{-2}_{\rm Pl}\Pi\right]a^2
\end{eqnarray}
These again can be shown to be equivalent to each other.

Therefore, for all components of the linearised Einstein equations that are relevant here, the two frames are physically identical -- not only do the equations take the same forms, but also they have the same physical content.

Note that in this subsubsection we used bold symbols to denote total quantities, including contributions from normal matter {\it and} the effective stress-energy tensor of the scalar field in the Jordan frame.
 
\subsubsection{Conservation equations}

Expressed in $k$ space, the perturbed continuity equations for photons in the Jordan and Einstein frames can be written as
\begin{eqnarray}\label{eq:Dgamma-k}
\left(\tilde{\Delta}^{(\gamma)}_{k}\right)' + \frac{4}{3}k\tilde{Z}_{k} -4\frac{\tilde{a}'}{\tilde{a}}\tilde{w}_{k} + k\tilde{v}^{(\gamma)}_{k} &=& 0,\nonumber\\
\left({\Delta}^{(\gamma)}_{k}\right)' + \frac{4}{3}k{Z}_{k} -4\frac{{a}'}{{a}}{w}_{k} + kv^{(\gamma)}_{k} &=& 0.
\end{eqnarray}
The Euler equations for photons can be written as
\begin{eqnarray}\label{eq:vgamma-k}
\left(\tilde{v}^{(\gamma)}_{k}\right)' - \frac{1}{3}k\tilde{\Delta}^{(\gamma)}_{k} + \frac{2}{3}k\tilde{\pi}^{(\gamma)}_{k} + \frac{4}{3}k\tilde{w}_{k} - \tilde{n}_{e}\tilde{\sigma}_{\rm T}\tilde{a}\left(\frac{4}{3}\tilde{v}^{(b)}_{k}-\tilde{v}^{(\gamma)}_{k}\right) &=& 0,\nonumber\\
\left({v}^{(\gamma)}_{k}\right)' - \frac{1}{3}k{\Delta}^{(\gamma)}_{k} + \frac{2}{3}k{\pi}^{(\gamma)}_{k} + \frac{4}{3}k{w}_{k} - {n}_{e}{\sigma}_{\rm T}{a}\left(\frac{4}{3}{v}^{(b)}_{k}-{v}^{(\gamma)}_{k}\right) &=& 0.
\end{eqnarray}
The perturbed continuity equations for cold dark matter become
\begin{eqnarray}
\left(\tilde{\Delta}^{(c)}_{k}\right)' + k\tilde{Z}_{k} - 3\frac{\tilde{a}'}{\tilde{a}}\tilde{w}_{k} + k\tilde{v}^{(c)}_{k} &=& 0,\nonumber\\
\left({\Delta}^{(c)}_{k}\right)' + k{Z}_{k} - 3\frac{{a}'}{{a}}{w}_{k} + kv^{(c)}_{k} &=& \frac{1}{2A}\xi'_{k} - \frac{A'}{2A^2}\xi_{k},
\end{eqnarray}
and the Euler equations are
\begin{eqnarray}
\left(\tilde{v}^{(c)}_{k}\right)' + \frac{\tilde{a}'}{\tilde{a}}\tilde{v}^{(c)}_{k} + k\tilde{w}_{k} &=& 0,\nonumber\\
\left({v}^{(c)}_{k}\right)' + \left[\frac{{a}'}{{a}}+\frac{A'}{2A}\right]{v}^{(c)}_{k} + k{w}_{k} &=& \frac{1}{2A}k\xi_{k}.
\end{eqnarray}
All these equations are equivalent between their Jordan frame and Einstein frame versions, as can be straightforwardly checked using the relations between the $k$-space quantities in the two frames given above. As in the real space case, we shall not present the relevant equations for baryons and massless neutrinos as they are similar.

Note that Eqs.~\eqref{eq:Dgamma-k} can be rewritten, in a form that more directly shows that they are `continuity' equations, as
\begin{eqnarray}
\left[\tilde{\Delta}^{(\gamma)}_{k} + 4\tilde{h}_{k}\right]' + k\tilde{v}^{(\gamma)}_{k} &=& 0,\nonumber\\
\left[{\Delta}^{(\gamma)}_{k} + 4{h}_{k}\right]' + k{v}^{(\gamma)}_{k} &=& 0,
\end{eqnarray}
where we have used $h'_{k}=\frac{1}{3}kZ_{k}-\frac{a'}{a}w_{k}$ and a similar relation in the Jordan-frame. Therefore, even though both $\Delta_{k}^{(\gamma)}$ and $h_{k}$ are frame-dependent, the value of their combination in the brackets are not because $\tilde{v}_{k}^{(\gamma)}=v_{k}^{(\gamma)}$. This is because $\left(\Delta^{(\gamma)}_{k}+4h_{k}\right)$ is the fractional perturbation of $\rho^{(\gamma)}a^4$, where $\rho^{(\gamma)}$ is the {\it local} photon energy density and $a$ is the {\it local} scale factor: the conformal transformation changes the size of a volume element and therefore the density in it, but it does not change the total energy inside the volume element. Doing the same for cold dark matter, we obtain
\begin{eqnarray}
\left[\tilde{\Delta}^{(c)}_{k} + 3\tilde{h}_{k}\right]' + k\tilde{v}^{(c)}_{k} &=& 0,\nonumber\\
\left[{\Delta}^{(\gamma)}_{k} + 3{h}_{k} - \frac{1}{2A}\xi_{k}\right]' + k{v}^{(\gamma)}_{k} &=& 0.
\end{eqnarray}
The equation in the Jordan frame can be understood as the mass conservation as in the case of photons, while the Einstein-frame equation looks different because of the variation of mass $m\propto A^{\frac{1}{2}}$ -- here what is conserved is not the mass in a volume element, $\rho^{(c)}a^3$, but the number of particles in it, which is proportional to $\rho^{(c)}A^{-\frac{1}{2}}a^3$.

\section{Frame-independence of cosmological observables}
\label{sect:observables}

In the previous sections we have explicitly checked the mathematical equivalence of the Einstein and conservation equations in the Jordan and Einstein frames. We have seen that although quantities such as the matter density perturbation, spatial curvature and gradient of the expansion scalar are different in these frames, and the matter contents in them are also not exactly the same, certain combinations of quantities are frame independent. To demonstrate the physical equivalence between the frames, we need to show that the quantities that are directly related to observables are frame independent. 

\subsection{The CMB power spectrum}

The CMB temperature map, whose anisotropy information is often presented in the form of its angular power spectrum $C(\ell)$, has been a primary cosmological observable, and can be used to simultaneously constrain all six cosmological parameters in the simple $\Lambda$CDM model. 

The CMB temperature anisotropies are primarily due to the inhomogeneities of photon densities at the time of last scattering, plus late-time secondary temperature fluctuations induced by the CMB photons falling in and climbing out of the potential wells created by the large-scale structures on their way to the observer. From the distribution function of photons, $f(E,e)$, where $E$ is photon energy and $e$ is the direction vector, the mean energy density can be written as
\begin{eqnarray}
\rho^{(\gamma)} &=& \int{\rm d}E{\rm d}\Omega{E^3}f(E,e) \propto {T}^4_{(\gamma)},
\end{eqnarray}
where $\Omega$ is the solid angle and $T_{(\gamma)}$ is the mean photon temperature. Hence, the direction-dependent CMB temperature fluctuation around the mean value is given by 
\begin{eqnarray}
\left[1+\delta_{\rm T}(e)\right]^4 = \frac{4\pi}{\rho^{(\gamma)}}\int{\rm d}EE^3f(E,e).
\end{eqnarray}
The $e$-dependence of the distribution function can be expanded using projected symmetric trace-free tensors as
\begin{eqnarray}
f &=& \sum_{\ell=0}^{\infty}F_{A_{\ell}}e^{A_{\ell}} = F + F_{\mu}e^{\mu} + F_{\mu\nu}e^{\mu}e^{\nu} + \cdots,
\end{eqnarray}
where $F$ is the unperturbed distribution function and $F_{\mu}, F_{\mu\nu}$ are first order quantities characterising the direction dependence. To linear order, this gives the following expansion \citep{3+1b,A_Lewis}
\begin{eqnarray}
\delta_{\rm T}(e) &=& \frac{1}{4}\sum_{\ell=1}^{\infty}\frac{(2\ell+1)!}{(-2)^{\ell}(\ell!)^2}I_{A_{\ell}}e^{A_{\ell}},
\end{eqnarray}
in which $I_{A_{\ell}}$ are projected symmetric trace-free energy-integrated multipoles of the distribution function
\begin{eqnarray}
I_{A_{\ell}} &\equiv& \frac{4\pi}{\rho^{(\gamma)}}\frac{(-2)^{\ell}(\ell!)^2}{(2\ell+1)!}\int^{\infty}_{0}{\rm d}EE^3F_{A_{\ell}}.
\end{eqnarray}
The collisional Boltzmann equation for photons can then be written order by order in $\ell$, which results in a hierarchy of coupled equations
\begin{eqnarray}\label{eq:boltzmann}
I'_{\ell,k} + k\left[\frac{\ell+1}{2\ell+1}I_{\ell+1,k}-\frac{\ell}{2\ell+1}I_{\ell-1,k}\right] + 4h'_{k}\delta^{0}_{\ell} + \frac{4}{3}kw_{k}\delta^{1}_{\ell} - \frac{8}{15}k\sigma_{k}\delta^{2}_{\ell} &=& -an_e\sigma_{\rm T}\left[I_{\ell,k}-\delta^{0}_{\ell}I_{0,k} - \frac{4}{3}\delta^{1}_{\ell}v^{(b)}_{k} - \frac{1}{10}\delta^{2}_{\ell}I_{2,k}\right],~~~~~~~~~
\end{eqnarray}
where $I_{\ell,k}$ is the $k$-space counterpart of $I_{A_{\ell}}$: $I_{A_{\ell}}=\sum_{k}I_{\ell,k}Q^{(k)}_{A_{\ell}}$, and $\delta^{0}_{\ell}$ etc. are Kronecker deltas. The right-hand side of Eq.~\eqref{eq:boltzmann} are collisional terms coming from Thomson scattering and we have neglected polarisation for the discussion here. The lowest three multipoles of $I_{\ell,k}$ ($\ell=0,1,2$) are respectively ${\Delta}_k^{(\gamma)}, v_k^{(\gamma)}, {\pi}_k^{(\gamma)}$, and one can check that the $\ell=0,1$ components of Eq.~\eqref{eq:boltzmann} are respectively Eqs.~\eqref{eq:Dgamma-k} and \eqref{eq:vgamma-k}. From the discussion in previous sections it follows that $\tilde{I}_{\ell,k}=I_{\ell,k}$ for all $\ell>0$, and hence the CMB temperature anisotropies should be the same in the two frames.

One can check this more explicitly. The solution to Eq.~\eqref{eq:boltzmann} can be written in the line-of-sight integral formula as \citep{A_Lewis}:
\begin{widetext}
\begin{eqnarray}\label{eq:los_int}
I_{\ell,k} (\eta_0) = 4 \int^{\eta_0} {\rm d}\eta{e}^{-\tau} \left(\left[k\sigma_{k}+\frac{3}{16}an_{e}\sigma_{\rm T}\pi^{(\gamma)}_{k}\right] \left[ \frac{1}{3}j_{\ell} + j^{\ast\ast}_{\ell} \right] + \left[an_{e}\sigma_{\rm T} v_k^{(b)}-kw_{k}\right]j^{\ast}_{\ell} + \left[\frac{1}{4}an_{e}\sigma_{\rm T} {\Delta}_k^{(\gamma)} - h_k'\right]j_{\ell}\right),~~~~~~
\end{eqnarray}
\end{widetext}
in which $j_{\ell}=j_{\ell}(x)=j_{\ell}(k(\eta_0-\eta))$ is the spherical Bessel function, $j^{\ast}_{\ell}(x)={\rm d}j_{\ell}(x)/{\rm d}x$,
and $\tau(\eta)$ is the optical depth defined by
\begin{eqnarray}
\tau(\eta) &\equiv& \int_{\eta}^{\eta_0}{\rm d}\eta a n_{e}\sigma_{\rm T},
\end{eqnarray}
where $\eta_0$ is the comoving time today. 

As discussed above, the conformal time $\eta$ is the same in the Jordan and Einstein frames, as well as $an_{e}\sigma_{\rm T}$ and therefore $\tau$. The spherical Bessel function is the radial part of the eigenfunction $Q^{(k)}$ of the comoving spatial d'Alembertian operator $a^2 \hat{\Box}$, and thus is the same in the two frames. Perturbed variables $\sigma_{k}, \pi^{(\gamma)}_{k}$ and $v^{(b)}_{k}$ are frame-independent too, and so we just need to check that the remaining terms in Eq.~\eqref{eq:los_int} are frame-independent -- this can be done by integrating by part the term involving $\Delta_{k}$ in Eq.~\eqref{eq:los_int}:
\begin{eqnarray}
\frac{1}{4}\int{\rm d}{\eta}{e}^{-{\tau}}{a}{n}_{e}{\sigma}_{\rm T}{\Delta}_k^{(\gamma)}j_{\ell} &=& \frac{1}{4}\int{\rm d}\eta\frac{{\rm d}{e}^{-\tau}}{{\rm d}\eta}\Delta^{(\gamma)}_{k}j_{\ell}(k(\eta_0-\eta)) = \frac{1}{4}\int{\rm d}\eta{e}^{-\tau}\left[k\Delta^{(\gamma)}_{k}j^{\ast}_{\ell} -\left(\Delta^{(\gamma)}_{k}\right)'j_{\ell}\right].
\end{eqnarray}
As combinations $\Delta_{k}-4w_{k}$ and $\Delta_{k}+4h_{k}$ are frame-independent (which can be checked using relations derived from previous sections), we conclude that $\tilde{I}_{\ell,k}=I_{\ell,k}$ and so the CMB temperature anisotropies are the same in the Jordan and Einstein frames.

\subsection{Other observables}

Apart from the primary CMB power spectrum, which depends on the Weyl potential $\Phi_{k}$ through the Sachs-Wolfe effect, there are other observables which are directly determined by $\Phi_{k}$. One is the integrated Sachs-Wolfe (ISW) effect, a secondary effect on the CMB temperature anisotropies caused by CMB photons gaining (or losing) energy by falling into and climbing out of time varying Weyl potentials, and which can be expressed as an integration over the time variation of $\Phi_k$:
\begin{eqnarray}
I^{\rm ISW}_{\ell,k} &=& 2\int^{\eta_0}_{\eta}{\rm d}\eta\Phi'_{k}j_{\ell}.
\end{eqnarray}
As part of the CMB temperature anisotropies, it is frame-independent as discussed in the previous subsection, and this can be seen directly as well, given that $\Phi_k$, $j_{\ell}$ and $\eta$ are the same in the two frames.

Another is gravitational lensing, which is the effect of the trajectories of photons from distant sources (such as galaxies or the last scattering surface) being deflected by the Weyl potential of foreground lenses (such as galaxy clusters, cosmic voids or more generally the intervening large-scale structure). This causes distortions of the images of the sources and amplifications of their magnitudes. The (unobserved) angular position of the source in the source plane, $\boldsymbol{\beta}$, is related to the observed angular position, $\boldsymbol{\theta}$, through
\begin{eqnarray}
\beta^{i} &=& \theta^{i} - \frac{2}{c^2}\int_0^{\chi_{\rm s}}{\rm d}\chi\frac{(\chi_s-\chi)\chi}{\chi_{\rm s}}\nabla^{\beta^i}\Phi_{\rm Weyl}\left(\chi,\boldsymbol{\beta}(\chi)\right),
\end{eqnarray}
where $\Phi_{\rm Weyl}$ is the Weyl potential (the real-space counterpart of $\Phi_k$), $\chi$ is the comoving distance, $\chi_{\rm s}$ is the comoving distance of the source and $i=(1,2)$ represent the two axes in the plane perpendicular to the line of sight. The comoving distance is given by $\chi=c\Delta\eta$ where $\Delta\eta$ is the conformal time needed for light to travel between the objects, and is frame-independent as $c$ and $\eta$. The Weyl potential is also frame-independent, so that it follows that the gravitational lensing calculated in the two frames are the same. 

The pre-recombination interaction between baryons, electrons and photons not only lead to the CMB temperature anisotropies, but is also responsible for a baryonic acoustic oscillation (BAO) length scale which is imprinted in the late-time distribution of matter, and which can be used as a standard ruler to measure cosmological distances. The BAO scale is given by the maximum distance traveled by sound waves before recombination, and has a comoving size of
\begin{eqnarray}
l_{\rm BAO} &=& \int_0^{\eta_{\rm rec}}c_{\rm s}{\rm d}\eta,
\end{eqnarray}
where $\eta_{\rm rec}$ is the conformal time of recombination and $c_{\rm s}$ is the speed of sound waves, given by
\begin{equation}
c_{\rm s} = \frac{1}{\sqrt{3 \left( 1+\frac{\rho^{(b)}}{\rho^{(\gamma)}} \right)}}.
\end{equation}
Because both ${\rho}^{(b)}$ and ${\rho}^{(\gamma)}$ transform in the same way in a conformal transformation, it follows that $c_{\rm s}$, and therefore $l_{\rm BAO}$, are frame-independent. The comoving angular diameter distance for BAO features at a given time $\eta$, $d_{\rm A}=l_{\rm BAO}/\Theta$, where $\Theta$ is the angle subtended by the BAO pattern, is equal to $\chi(\eta)$ for a flat space, and is frame-independent.

While the relation between the comoving distance and conformal time, $\chi(\eta)$, is frame-independent, the same does not apply if other `time' variables are used. For example, the scale factor, which is often used as a time variable in cosmology, is different in the two frames: $\tilde{a}=\sqrt{A}a$. From the equations
\begin{eqnarray}
\tilde{a}_0 &=& \int^{\eta_0}_0{\rm d}\eta\frac{{\rm d}\tilde{a}}{{\rm d}\eta},\nonumber\\
a_0 &=& \int^{\eta_0}_0{\rm d}\eta\frac{{\rm d}a}{{\rm d}\eta}\ =\ \int^{\eta_0}_0{\rm d}\eta\frac{{\rm d}}{{\rm d}\eta}\left(\tilde{a}A^{-\frac{1}{2}}\right),
\end{eqnarray}
it can be seen that $\tilde{a}_0=1$ for today in the Jordan frame corresponds to $a_0\neq1$ in the Einstein frame, where today is characterised by $a_0=A_0^{-\frac{1}{2}}$. This means that the comoving-distance-scale-factor relations are different in the frames. The physical times are also frame-dependent as ${\rm d}\tilde{t}=\sqrt{A}{\rm d}t$. Because time measurements require the use of atomic transitions, which are not affected by the scalar field in the Jordan frame where matter is minimally coupled, we consider $\tilde{t}$ as the physical time, and it is convenient to define $\tilde{a}_0=\tilde{a}(\tilde{t}_0)=1$. 

In the Jordan frame, cosmological redshift is given as usual: $\tilde{z}=1/\tilde{a}-1$. In the Einstein frame it is a bit more complicated: due to the time evolution of particle masses, including the electron mass, in this frame, the frequency of an atomic transition as measured in the past (e.g., when the conformal time was $\eta$), $\nu(\eta)$, is not the same as the frequency of the same atomic transition measured in our labs today, $\nu_0$, but the two are related by $\nu(\eta)=\sqrt{A(\eta_0)/A(\eta)}\nu_0$ since $\nu\propto m=\tilde{m}A^{-\frac{1}{2}}$ \citep{Veiled_GR}. The total photon redshifting including this contribution is then $1+z\equiv\nu(\eta)/\nu_0=[A(\eta_0)^{\frac{1}{2}}a_0]/[A(\eta)^{\frac{1}{2}}a(\eta)] = \tilde{a}_0/\tilde{a}=1/\tilde{a}=1+\tilde{z}$. Therefore, redshift is a frame-independent quantity. The luminosity distance of an object a photon emitted by which at time $\eta$ is received by an observer today is given by $d_{\rm L}=(1+z)\chi(\eta)$ where $\chi(\eta)$ is the comoving distance to $\eta$, and the $(1+z)$ factor comes from photon energy redshifting and time dilation, both of which are affected by the $A$-dependence of frequency -- with the properly defined frame-independent redshift, the luminosity-distance-redshift relation $d_{\rm L}(z)$ is the same in the two frames as well.

Finally, another commonly used cosmological observable is the two point correlation function of the matter overdensity field or its tracers, $\xi(|{\bf r}|)=\langle\Delta({\bf x})\Delta({\bf x}+{\bf r})\rangle$, or the matter power spectrum $P(k)$ given by $\langle\Delta_k({\bf k})\Delta_{k}({\bf k}')\rangle=(2\pi)^3\delta({\bf k}-{\bf k}')P(|{\bf k}|)$. As we have seen in Eq.~\eqref{eq:Dk-transform}, $\Delta_{k}$ is frame-dependent, which means that $P(k)$ depends on whether we are in the Einstein or the Jordan frames. Note that $P(k)$ is generally gauge dependent, but the effect of using different gauges is small on small scales.

\subsection{A numerical example}

\begin{figure*}
\includegraphics[width=18cm]{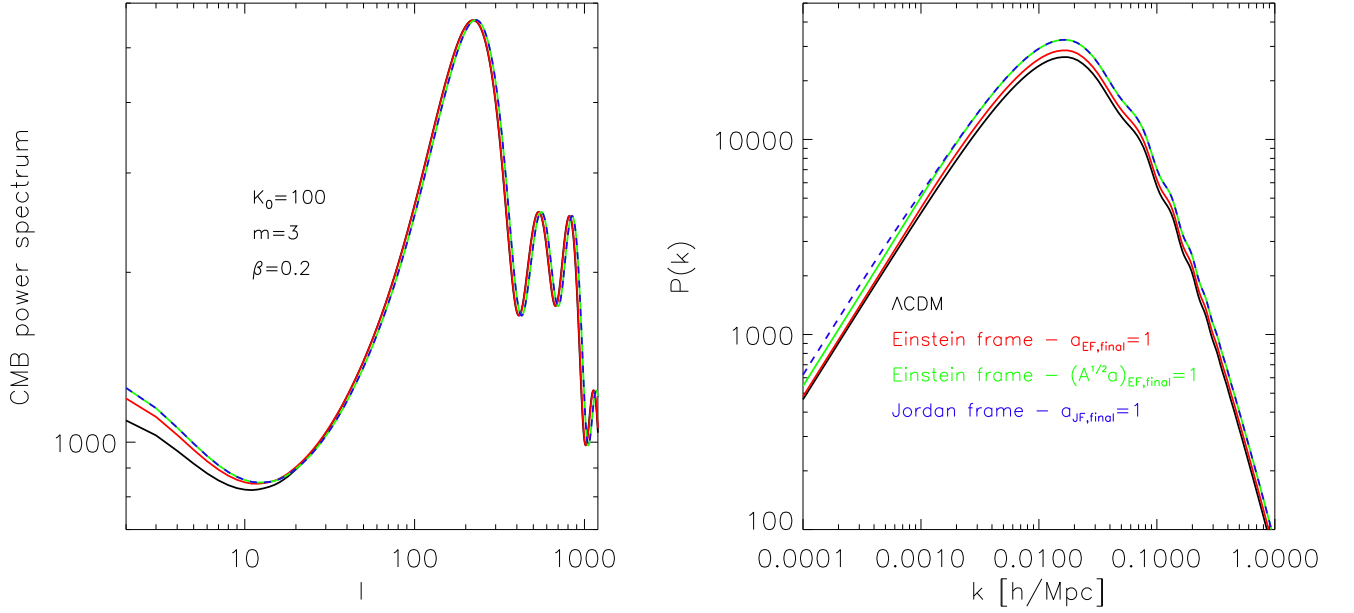}
\caption{(Colour Online) {\it Left Panel}: the CMB temperature power spectrum $C(l)$ in the different models/calculations. The black solid line is for the $\Lambda$CDM model for comparison, the blue dashed and green solid lines are the results computed from the Jordan-frame and Einstein-frame quantities respectively, and the red solid line is computed in the Einstein frame but with the calculation stopping at $a=1$ rather than $\sqrt{A}a=1$. {\it Right Panel}: the same as the left panel but for the matter power spectra $P(k)$.}
\label{fig:lin}
\end{figure*}

From the discussion and comparisons above, it is apparent that the Jordan frame has the disadvantage of having complicated expressions. Taking Eqs.~\eqref{eq:weyl_in_k} as example: the tilded dynamical quantities in the Jordan-frame version of these equations, given in Eqs.~\eqref{eq:Xkexp}, \eqref{eq:qkexp} and \eqref{eq:pikexp} respectively, are lengthy and in the end they cancel each other in a combination. Therefore, the Einstein frame is computationally more convenient in practice.

It is often said that the calculation can be done in either of the frames, and physical observables should not depend on which frame is used. While this is true, there is a subtlety here -- the two frames have the same redshift, but not the same values of the scale factor, i.e., $\tilde{a}\neq{a}$. To obtain observables today, such as the CMB power spectrum, the linear perturbation equations usually need to be integrated up to $z=0$ or $\tilde{a}=1$, and if the calculation is carried out in the Einstein frame the integration should be stopped at $a\neq1$. Therefore, a recipe for linear perturbation calculation is to compute the background quantities using $\tilde{a}$ -- which has the advantage of being more directly related to redshifts -- and the perturbation evolution using the conformal time $\eta$ -- which is the same in both frames. There is no need to derive the Jordan-frame background equations, even though they should be fairly simple, and in practice one can do everything in the Einstein frame: $\tilde{a}$ is given by $\sqrt{A}a$. One example to show why it is more convenient to use $\tilde{a}$ instead of $a$ is the conformal time today, for which we need to integrate ${\rm d}\eta/{\rm d}\tilde{a}$ until $\tilde{a}=1$, while if we use $a$ the integration should stop at $a=A^{-1/2}$ which is model dependent\footnote{In many places the codes are default to integrate to $z=1/\tilde{a}-1=0$, e.g., the function {\tt dtauda} in {\sc camb}. If we integrate ${\rm d}\eta/{\rm d}a$ these places may all need to be changed.}.

As an numerical example, we consider the K-mouflage model \cite{kmouflage1,kmouflage2} studied in \cite{K-mouflage}. In this model, the total action is given by Eq.~\eqref{action EF} with 
\begin{eqnarray}
A(\phi) &=& \exp(2\beta{M}_{\rm Pl}^{-1}\phi),\\
\mathcal{L}_{\phi} &=& -M^4K(\sigma),
\end{eqnarray}
where $\beta$ is a dimensionless model parameter characterising the coupling strength of the scalar field with matter in the Einstein frame, $M$ is a model parameter of mass dimension that will be fixed given the fractional energy density of the scalar field today,
\begin{eqnarray}
\sigma &\equiv& \frac{1}{2}M_{\rm Pl}^{-4}(\nabla\phi)^2,\\
K &=& -1+\sigma+K_0\sigma^m,
\end{eqnarray}
where $K_0, m$ are two other dimensionless parameters. This model has been described in details in the above references, and as we only use it as an example to illustrate our numerical implementation, we shall keep things simple by only presenting the above equations. Note that $\sigma$ here has no subscript $_k$, to be distinguished from the shear $\sigma_k$.

We have implemented this model in the publicly available linear Boltzmann code {\sc camb} \cite{camb}, using both the Einstein and the Jordan frames, and some numerical results are shown in Figure~\ref{fig:lin}. In the left panel we have plotted the CMB temperature spectra for the K-mouflage model (see legends for model parameters) as the coloured curves, and the corresponding $\Lambda$CDM model\footnote{This is the $\Lambda$CDM model whose present-day density parameter $\Omega_\Lambda$ is equal to the current density parameter of the scalar field, $\Omega_{\phi0}$, in the K-mouflage model, and all the non-Kmouflage model parameters are the same in the two models.} as the black solid line. The blue dashed and green solid curves are obtained respectively by integrating Jordan- and Einstein-frame perturbation quantities to $\sqrt{A}a=1$ and $\tilde{a}=1$ (in this particular model we find that $a=1.0835$ at $\tilde{a}=1$), and they are identical as expected from the discussion above. The red solid line, in contrast, is obtained by integrating the Einstein-frame variables to $a=1$: here the background expansion history is incorrect (therefore the shift of CMB peaks due to wrong distances) and there is less time for the evolution than in the correct calculation (hence a weaker integrated Sachs-Wolfe effect).

In the right panel of Figure \ref{fig:lin}, we show the linear matter power spectra for the same models/calculations. Again, because the red curve stops at $a=1$ rather than $a=1.0835$, there has been less time for the growth of matter density perturbations, which results in a smaller $P(k)$ than the correct prediction. The blue dashed and green solid lines are identical on small scales, while on very large scales they show mild difference. The density contrast, and therefore $P(k)$, is gauge-dependent, and this difference is expected unless one uses the gauge in which the scalar field is homogeneous ($\xi_k=0$).

\section{Discussions and conclusions}
\label{sect:discussion}

In this paper, we have studied some aspects of the physics of the Scalar-Tensor theory in two conformally related frames, the so-called Einstein and Jordan frames. Many debates occurred in the community about their physical equivalence, and this work aims to confirm the idea that the two frames are physically equivalent, namely results of cosmological observations are the same in both frames, given that they are basically the same action expressed in different forms using field redefinition.

We have done this by a detailed comparison of the equations in the two frames at different levels. Starting from the original action written in the Einstein frame, we re-express it in the Jordan frame, and derive the Einstein and Klein-Gordon equations in the two frames, before checking that they are mathematically equivalent. This means that one can derive an equation of motion in one frame simply by starting from its counterpart in the other frame. In other words, working in parallel and independently in both frames, or working in a given frame and then moving to the other, are two equivalent approaches to study these theories.

% which can be non equivalent at a quantum level, arXiv:1010.4536 : quantization and change of frame do not commute

We have then focused in Section \ref{sect:pert_eqns} on the links between physical quantities (that is, dynamical and kinematic quantities) in the two frames, and used these relations to show the equivalence of key linearised perturbed equations. Many physical quantities (such as the scale factor, the 4-acceleration or the energy density of a given matter specie for instance) have different expressions in the two frames, which could lead to the conclusion that the they are not physically equivalent. However, the crucial point to notice is that these quantities are not directly measurable by an observer. The cosmological observables arise from combinations of these quantities, and Section \ref{sect:observables} shows that these combinations are generally frame independent. Hence, computations can be done in either frame without changing physical conclusions.

Physics in the Einstein frame is described by GR, to which a new specie of matter is added -- a scalar field which will interact will usual species of matter. Hence, in the Einstein frame, Einstein equations have the same form as in GR, while conservation equations of different matter species have different forms from the standard model, since matter exchange energy and momentum with the scalar field. For cosmological perturbations, working in the Einstein frame has the advantage of substantially simplified field equations. In the Jordan frame, gravity is no longer described by GR, as the scalar-field is now non-minimally coupled to the gravitational part of the action so that the Einstein equations are more complicated. The corrections to the standard GR equations can be treated as an effective fluid which contributes terms to the total (effective) stress energy tensor. In the example of the constraint equation for the Weyl potential, such complicated additional terms cancel exactly, leaving the result unchanged from the much simpler Einstein-frame calculation. 

Because the two frames are related by a conformal transformation, they share the same conformal time ($\eta$), which means that calculations using $\eta$ as the time variable are not affected. However, the scale factor takes different values in the two frames, with the Jordan-frame scale factor $\tilde{a}$ considered as the physical one because matter particles follow their geodesics in this frame. The redshift, $\tilde{z}=z=\frac{1}{\tilde{a}}-1$ is a physical observable that is agreed by both frames, even though $z\neq1/a-1$ in the Einstein frame. 
% On the other hand, computations involving redshift are easier to carry out in the JF, as in this frame we have the well-known relation $1+ \tilde z = \frac{\tilde a_0}{\tilde a}$, which is wrong in the EF. 
Instead, in the Einstein frame we have $z=1/\left(\sqrt{A}a\right)-1$; when working in this frame, the results of observables, such as the CMB power spectrum, may not automatically be the same as from the Jordan-frame calculation, and care needs to be taken to ensure that integrations end at the correct time, as $a_0=A_0^{-\frac{1}{2}}$ rather than $a_0=1$ represents the present day.
% If one chooses the useful convention that today $\tilde a_0 = 1$ in the JF and decide to work in the Einstein frame, we must take care that we will no longer have $a_0=1$ in this frame.\\

Before closing the paper, we would like to mention that the cosmological equivalence studied here is at the classical level. There have been interesting discussions at the quantum level, e.g., \cite{quantum effects}, and these are beyond the scope of this work.

Another interesting point which could motivate further works is the case when the two metrics are related by more complicated relations than the conformal one \eqref{conformal transformation} studied all along this paper. For example, it is known that \cite{Bekenstein} the most general relation linking the metrics and a scalar field $\phi$, compatible with causality and the weak equivalence principle, is a disformal transformation $\tilde g_{\mu \nu} = A(\phi) g_{\mu \nu} + B(\phi) \nabla_{\mu} \phi \nabla_{\nu} \phi$. As in the conformal case, $(\mathcal{M}, \tilde g_{\mu \nu})$ defines the Jordan frame, while $(\mathcal{M}, g_{\mu \nu})$ defines what we call the Einstein frame. While a purely conformal transformation is merely a rescaling of the metric, a disformal transformation contains both a conformal rescaling of the metric and a distortion of it. Such transformations have been considered in various circumstances, such as varying speed of light cosmologies \cite{varying c}, theories of massive gravity \cite{massive gravity} or in the description of branes embedded in a higher dimensional space \cite{branes}, and the physical equivalence of the Einstein and Jordan frames in this general case could be particularly interesting to study, see, e.g., \cite{disformal_1,disformal_2,disformal_3,disformal_4,disformal_5}.

\acknowledgments
FR is supported by Ecole Normale Sup\'{e}rieure Paris-Saclay, and thanks for the host of the Institute for Computational Cosmology (ICC) at Durham University where the work described here was carried out. BL is supported by the European Research Council (ERC-StG-716532-PUNCA), the ICC's STFC Consolidated Grants (ST/P000541/1, ST/L00075X/1) and Durham University.

\appendix

\section{Useful perturbation relations}

In this appendix we present some useful relations that hold to first-order in perturbations, which are useful for derivations and checks of the perturbed equations.

We start from the relation between the second-order covariant derivatives in the two frames $\tilde{\nabla}_{\mu}\tilde{\nabla}^{\nu}\psi$ and ${\nabla}_{\mu}{\nabla}^{\nu}\psi$, where $\psi$ is a general scalar quantity
\begin{widetext}
\begin{eqnarray}
\tilde{\nabla}_{\mu}\tilde{\nabla}^{\nu}\psi &=& \frac{1}{A}{\nabla}_{\mu}{\nabla}^{\nu}\psi - \frac{1}{2A}\left(\nabla_{\mu}A\nabla^{\nu}\psi+\nabla_{\mu}\psi\nabla^{\nu}A-\delta^{\mu}_{\nu}\nabla^{\lambda}A\nabla_{\lambda}\psi\right).
\end{eqnarray}
\end{widetext}
Using the decomposition of these covariant derivatives in the two frames:
\begin{widetext}
\begin{eqnarray}
\nabla_{\mu}\nabla^{\nu}\psi &=& \hat{\nabla}_{\langle\mu}\hat{\nabla}^{\nu\rangle}\psi + \frac{1}{3}h_{\mu}^{\ \nu}\hat{\Box}\psi + u_{\mu}u^{\nu}\ddot{\psi} - 2u_{(\mu} \hat{\nabla}^{\nu)}\dot{\psi} + \frac{2}{3}\theta{u}_{(\mu}\hat{\nabla}^{\nu)}\psi - \sigma_{\mu}^{\ \nu}\dot{\psi} - \varpi_{\mu}^{\ \nu}\dot{\psi} - \frac{1}{3}\theta{h}_{\mu}^{\ \nu}\dot{\psi},\\
\tilde{\nabla}_{\mu}\tilde{\nabla}^{\nu}\psi &=& \hat{\tilde{\nabla}}_{\langle\mu}\hat{\tilde{\nabla}}^{\nu\rangle}\psi + \frac{1}{3}\tilde{h}_{\mu}^{\ \nu}\hat{\tilde{\Box}}\psi + \tilde{u}_{\mu}\tilde{u}^{\nu}\mathring{\mathring{\psi}} - 2\tilde{u}_{(\mu} \hat{\tilde{\nabla}}^{\nu)}\mathring{\psi} + \frac{2}{3}\tilde{\theta}\tilde{u}_{(\mu}\hat{\tilde{\nabla}}^{\nu)}\psi - \tilde{\sigma}_{\mu}^{\ \nu}\mathring{\psi} - \tilde{\varpi}_{\mu}^{\ \nu}\mathring{\psi} - \frac{1}{3}\tilde{\theta}\tilde{h}_{\mu}^{\ \nu}\mathring{\psi},
\end{eqnarray}
\end{widetext}
and that (to first order)
\begin{eqnarray}
\nabla_{\mu}A\nabla^{\nu}\psi &=& \dot{A}\dot{\psi}u_{\mu}u^{\nu} - u_{\mu}\dot{A}\hat{\nabla}^{\nu}\psi - u^{\nu}\dot{\psi}\hat{\nabla}_{\mu}A,\nonumber\\
\nabla_{\mu}\psi\nabla^{\nu}A &=& \dot{A}\dot{\psi}u_{\mu}u^{\nu} - u_{\mu}\dot{\psi}\hat{\nabla}^{\nu}A - u^{\nu}\dot{A}\hat{\nabla}_{\mu}\psi,\nonumber\\
\nabla^{\lambda}A\nabla_{\lambda}\psi &=& -\dot{A}\dot{\psi},\nonumber
\end{eqnarray}
it can be found
\begin{eqnarray}
\tilde{\theta} &=& \frac{1}{A^{\frac{1}{2}}}\left[\theta+\frac{3\dot{A}}{2A}\right],\\
\tilde{\sigma}_{\mu}^{\ \nu} &=& \frac{1}{A^{\frac{1}{2}}}\sigma_{\mu}^{\ \nu},\\
\tilde{\varpi}_{\mu}^{\ \nu} &=& \frac{1}{A^{\frac{1}{2}}}\varpi_{\mu}^{\ \nu},\\
\hat{\tilde{\Box}}\psi &=& \frac{1}{A}\hat{\Box}\psi,\\
\hat{\tilde{\nabla}}_{\langle\mu}\hat{\tilde{\nabla}}^{\nu\rangle}\psi &=& \frac{1}{A}\hat{\nabla}_{\langle\mu}\hat{\nabla}^{\nu\rangle}\psi,\\
\hat{\tilde{\nabla}}_{\mu}\psi &=& \hat{\nabla}_{\mu}\psi,\\
\hat{\tilde{\nabla}}^{\mu}\psi &=& \frac{1}{A}\hat{\nabla}^{\mu}\psi,\\
\hat{\tilde{\nabla}}_{\mu}\mathring{\psi} &=& \frac{1}{A^{\frac{1}{2}}}\left[\hat{\nabla}_{\mu}\dot{\psi}-\frac{\dot{A}}{2A}\hat{\nabla}_{\mu}\psi\right],\\
\hat{\tilde{\nabla}}^{\mu}\mathring{\psi} &=& \frac{1}{A^{\frac{3}{2}}}\left[\hat{\nabla}^{\mu}\dot{\psi}-\frac{\dot{A}}{2A}\hat{\nabla}^{\mu}\psi\right].
\end{eqnarray}

In the $3+1$ formalism, time and covariant spatial derivatives do not commute, but they satisfy the following relation which is useful in calculations:
\begin{equation}\label{C1}
\hat{\nabla}_{\mu} \dot \psi = (\hat{\nabla}_{\mu} \psi)^{\cdot} + \frac{1}{3} \theta \hat{\nabla}_{\mu} \psi - \dot u_{\mu} \dot \psi.
\end{equation}
A similar relation exists for the Jordan frame quantities.

\section{Equivalence of the Klein-Gordon equations in the Jordan and Einstein frames}
\label{sect:append_b}

Let's start with the Klein-Gordon equation in the Jordan frame
\begin{equation}\label{KG eq JF 2}
\tilde \nabla_{\mu} \left[ \frac{\partial \tilde{\mathcal L}_{\tilde \phi} (\tilde \phi, (\tilde \nabla \tilde \phi)^2)}{\partial (\tilde \nabla_{\mu} \tilde \phi)} \right] = \frac{\partial \tilde {\mathcal L}_{\tilde \phi}}{\partial \tilde \phi} \\ + \frac{1}{2} \frac{{\rm d}\ln A}{{\rm d}\tilde \phi} \left(\tilde T^{(m)} + \tilde T^{(\tilde \phi)} \right).
\end{equation}

To show that this equation is equivalent to its Einstein-frame counterpart, we first slightly rewrite Eq.~\eqref{covariant derivative of derivative of Lagrangian} as
\begin{equation}\label{covariant derivative of derivative of Lagrangian2}
\tilde \nabla_{\mu} \left[ \frac{\partial \tilde{\mathcal L}_{\tilde \phi}}{\partial (\tilde \nabla_{\mu} \tilde \phi)} \right] = \frac{1}{A^{\frac{3}{2}}} \nabla_{\mu} \left[ \frac{\partial {\mathcal L}_{\phi}}{\partial (\nabla_{\mu} \phi)} \right] + \frac{1}{2A^{\frac{3}{2}}}\frac{{\rm d}\ln A}{{\rm d}\phi}\frac{\partial\mathcal{L}_{\phi}}{\partial\sigma}\left(\nabla\phi\right)^2,
\end{equation}
where we have used $\frac{\partial\mathcal{L}_{\phi}}{\partial(\nabla_{\mu}\phi)}\nabla_{\mu}\phi=\frac{\partial\mathcal{L}_{\phi}}{\partial\sigma}\left(\nabla\phi\right)^2$ where $\sigma\equiv\frac{1}{2}\left(\nabla\phi\right)^2$.

We will also use the relations $\partial{\tilde{\phi}}/\partial\phi=A^{-\frac{1}{2}}$, $\tilde{\mathcal{L}}_{\tilde{\phi}}=A^{-2}\mathcal{L}_{\phi}$, $\tilde{T}^{(i)}=A^{-2}T^{(i)}$ and $T^{(\phi)}=4\mathcal{L}_{\phi}-\frac{\partial\mathcal{L}_{\phi}}{\partial\sigma}\left(\nabla\phi\right)^2$, where the last one comes from 
\begin{eqnarray}
T_{\mu\nu}^{(\phi)} = -\frac{2}{\sqrt{-g}}\frac{\delta\left(\sqrt{-g}\mathcal{L}_{\phi}\right)}{\delta{g}^{\mu\nu}} = \mathcal{L}_{\phi}g_{\mu\nu} - \frac{\partial{\mathcal{L}_{\phi}}}{\partial\sigma}\nabla_{\mu}\phi\nabla_{\nu}\phi.
\end{eqnarray}

Finally, for a $\mathcal{L}_{\phi}$ that contains general functions of $\sigma=\frac{1}{2}\left(\nabla\phi\right)^2$, to ensure the correct dimension, the Lagrangian density can be written for example as 
\begin{eqnarray}
\mathcal{L}_{\phi}\left(\phi,\sigma\right) &=& M_\ast^{4}K\left(M_\ast^{-4}{\sigma}\right) - V(\phi),
\end{eqnarray} 
where $M_\ast$ is a constant of mass dimension and $K(\cdots)$ is a dimensionless function. The Jordan-frame counterpart has the form 
\begin{eqnarray}
\tilde{\mathcal{L}}_{\tilde{\phi}}\left(\tilde{\phi},\tilde{\sigma}\right) &=& \left(A^{-\frac{1}{2}}M_\ast\right)^{4}K\left[\left(A^{-\frac{1}{2}}M_\ast\right)^{-4}{\tilde{\sigma}}\right] - \tilde{V}\left(\tilde{\phi}\right),
\end{eqnarray}
where $\tilde{\sigma}=\frac{1}{2}\left(\tilde{\nabla}\tilde{\phi}\right)^2=A^{-2}\sigma$. Therefore, we have
\begin{eqnarray}
\frac{\partial\tilde{\mathcal{L}}_{\tilde{\phi}}}{\partial\tilde{\phi}} &=& -\frac{2}{A^{\frac{3}{2}}}\frac{{\rm d}\ln A}{{\rm d}\phi}\mathcal{L}_{\phi} + \frac{1}{A^{\frac{3}{2}}}\frac{\partial{\mathcal{L}}_{{\phi}}}{\partial{\phi}} + \frac{1}{A^{\frac{3}{2}}}\frac{\partial{K}}{\partial\sigma}\frac{{\rm d}\ln A}{{\rm d}\phi}\left(\nabla\phi\right)^2.
\end{eqnarray}
Note that $\partial{K}/\partial\sigma=\partial\mathcal{L}_{\phi}/\partial\sigma$. 
% In the simplest case of a quintessence model, with 
% \begin{eqnarray}
% \mathcal{L}_{\phi} = -\sigma-V(\phi) = -\frac{1}{2}g^{\mu\nu}\nabla_{\mu}\phi\nabla_{\nu}\phi - V(\phi),
% \end{eqnarray}
% we have $\partial{K}/\partial\sigma=-1$ and so this simplifies to
% \begin{eqnarray}
% \frac{\partial\tilde{\mathcal{L}}_{\tilde{\phi}}}{\partial\tilde{\phi}} &=& \frac{2}{A^{\frac{3}{2}}}\frac{{\rm d}\ln A}{{\rm d}\phi}V(\phi) + \frac{1}{A^{\frac{3}{2}}}\frac{\partial{\mathcal{L}}_{{\phi}}}{\partial{\phi}}.
% \end{eqnarray}

Using the above relations, it is straightforward to check that Eq.~\eqref{KG eq JF 2} can be rewritten as
\begin{equation}\label{KG eq EF}
\nabla_{\mu} \left[ \frac{\partial \mathcal{L}_{\phi} (\phi, (\nabla \phi)^2)}{\partial(\nabla_{\mu} \phi)} \right] = \frac{\partial{\mathcal{L}}_{\phi}}{\partial{\phi}} + \frac{1}{2} \frac{{\rm d}\ln A}{{\rm d}\phi} T^{(m)},
\end{equation}
which is the Klein-Gordon equation in the Einstein frame.

% \begin{equation}\label{dot_dalembertian}
% \hat{\Box} \dot \psi = (\hat{\Box} \psi)^{\cdot} + \frac{2}{3} \theta \hat{\Box} \psi - \dot \psi \hat{\nabla}_{\alpha} \dot u^{\alpha}
% \end{equation}

% The quantities $\nabla_{\mu}\nabla_{\nu}\psi$ and $\nabla_{\mu} \psi \nabla_{\nu} \psi$ can be respectively split as :
% \begin{multline}\label{A1}
% \nabla_{\mu} \nabla_{\nu} \psi = \hat{\nabla}_{<\mu} \hat{\nabla}_{\nu>} \psi + \frac{1}{3} h_{\mu \nu} \hat{\Box} \psi + u_{\mu} u_{\nu} \ddot \psi \\ - 2 u_{(\mu} \hat{\nabla}_{\nu)} \dot \psi + \frac{2}{3} \theta u_{(\mu} \hat{\nabla}_{\nu)} \psi - \sigma_{\mu \nu} \dot \psi - \frac{1}{3} \theta h_{\mu \nu} \dot \psi
% \end{multline}

% and

% \begin{equation}\label{A2}
% \nabla_{\mu} \psi \nabla_{\nu} \psi = u_{\mu} u_{\nu} \dot \psi^2 - 2 \dot \psi u_{(\mu} \hat{\nabla}_{\nu)} \psi
% \end{equation}

% It is then straightforward to deduce from \eqref{A1} that :

% \begin{equation}
% u^{\mu} u^{\nu} \nabla_{\mu} \nabla_{\nu} \psi = \ddot \psi
% \end{equation}
% and
% \begin{equation}\label{dalembertian}
% \Box \psi = \hat{\Box} \psi - \ddot \psi - \theta \dot \psi
% \end{equation}

% and from \eqref{A2} that :

% \begin{equation}
% u^{\mu} u^{\nu} \nabla_{\mu} \psi \nabla_{\nu} \psi = \dot \psi^2
% \end{equation}
% and
% \begin{equation}
% (\nabla \psi)^2 = - \dot \psi^2
% \end{equation}

\end{document}